\documentclass[12pt]{iopart}

\usepackage{bm}
\usepackage{amsopn}
\usepackage{amssymb}
\usepackage{setstack}
\usepackage{feynmp}
\RequirePackage{ifpdf}

\ifpdf
\else % ordinary latex seems to include these as eps files without a problem
\fi

\eqnobysec

\newcommand{\ep}{\epsilon}

\newcommand{\lle}{\left<}
\newcommand{\rgr}{\right>}
\newcommand{\lb}{\left|}
\newcommand{\rb}{\right|}

\def\bk{{\bf k}}
\def\bZ{{\bf Z}}
\newcommand{\fnl}{\ensuremath{f_{\mathrm{NL}}}}
\newcommand{\tnl}{\ensuremath{\tau_{\mathrm{NL}}}}
\newcommand{\gnl}{\ensuremath{g_{\mathrm{NL}}}}
\newcommand{\Planck}{M_{\mathrm{P}}}
\newcommand{\R}{\mathcal{R}}
\newcommand{\Hint}{H_{\mathrm{I}}}
\newcommand{\Torder}{\mathrm{T}}

\renewcommand{\d}{\mathrm{d}}
\newcommand{\vect}[1]{\bm{\mathrm{{#1}}}}
\renewcommand{\e}[1]{\mathrm{e}^{{#1}}}
\newcommand{\im}{\mathrm{i}}
\newcommand{\dpp}{\varphi}

\renewcommand{\leq}{\leqslant}

\DeclareMathOperator{\RePart}{Re}
\DeclareMathOperator{\ImPart}{Im}
\renewcommand{\Re}{\RePart}
\renewcommand{\Im}{\ImPart}

\newcommand{\flabel}[1]{{\small(\textsf{{#1}})}}

\newcommand{\ie}{\emph{i.e.}}

\begin{document}
\begin{fmffile}{diags}

\title{Inflationary trispectrum from graviton exchange}

\author{David Seery$^1$, Martin S. Sloth$^2$, Filippo Vernizzi$^{3,4}$}

\address{$^1$ Department of Applied Mathematics and Theoretical Physics, \\
Wilberforce Road, Cambridge, CB3 0WA, United Kingdom}
\vspace{2mm}
\address{$^2$Department of Physics and Astronomy, University of Aarhus \\
Ny Munkegade, DK-8000 Aarhus C, Denmark}
\vspace{2mm}
\address{$^3$CEA, IPhT, 91191 Gif-sur-Yvette c\'edex,
France\footnote{Permanent address}\\
CNRS, URA-2306, 91191 Gif-sur-Yvette c\'edex, France}
\vspace{2mm}
\address{$^4$Abdus Salam ICTP, Strada Costiera 11,
34014 Trieste, Italy}

\ead{\mailto{djs61@cam.ac.uk}, \mailto{sloth@phys.au.dk},
 \mailto{filippo.vernizzi@cea.fr}}

\begin{abstract}
  We compute the connected four-point correlation function of the
  primordial curvature perturbation generated during
  inflation with standard kinetic terms, where the correlation is
  established via exchange of a graviton between two pairs of scalar
  fluctuations.
  Any such correlation yields a contribution to the scalar trispectrum
  of the order of the tensor to scalar ratio $r$. This contribution
  is numerically one order of magnitude larger than the one
  previously calculated on the basis of scalar perturbations
  interacting at a point and
  satisfies a simple relation in the
  limit where the momentum of the graviton which is exchanged
  becomes much smaller than the external momenta.
  We conclude that the total non-linearity parameter generated
  by single-field models of slow-roll inflation is at maximum $|\tnl| \sim r$.
\end{abstract}

\maketitle

%%%%%%%%%%%%%%%%%%%%%%%%%%%%%%%%%%%%%%%%%%%%%%%%%%%%%%%%%%%%%%%%%%%%%%%%%%%%
%%%%%%%%%%%%%%%%%%%%%%%%%%%%%%%%%%%%%%%%%%%%%%%%%%%%%%%%%%%%%%%%%%%%%%%%%%%%

\section{Introduction}
\label{sec:introduction}

%%%%%%%%%%%%%%%%%%%%%%%%%%%%%%%%%%%%%%%%%%%%%%%%%%%%%%%%%%%%%%%%%%%%%%%%%%%%
%%%%%%%%%%%%%%%%%%%%%%%%%%%%%%%%%%%%%%%%%%%%%%%%%%%%%%%%%%%%%%%%%%%%%%%%%%%%

Since the release of the first-year \emph{Wilkinson Microwave
Anisotropy Probe} (WMAP) satellite data, the availability of
high-resolution all-sky maps of the Cosmic Microwave Background
(CMB) anisotropy has revolutionized cosmology as a
quantitative science. Together with
large galaxy surveys, the ever-improving accuracy of the microwave
background data has yielded a great deal of precise information
concerning conditions in the very early universe. Over the next
several years a new generation of CMB instruments will return yet
more data, which will fix the parameters of the standard model
with even greater precision and may point the way to the discovery
of new physics.

Despite its role as the dominant carrier of information from
the earliest times, the statistical properties of the CMB temperature
anisotropy are remarkably simple: it
is known to be statistically Gaussian to very high
precision, greater than $0.1$\%.
In recent years, however, the possibility that small non-Gaussian
features might be present in these fluctuations has received
increasing interest from both cosmologists and particle
physicists. This is because primordial non-Gaussianities are
one of our best
probes of whatever complex physics was operating in the
very early universe. In particular, non-Gaussian features
discriminate sharply between the minimal implementation of
inflation---driven by a single scalar field rolling down its
potential under the influence of Hubble friction, known as
``slow-roll'' dynamics---and more complicated variants involving
extra fields or dynamics of greater subtlety.

Which observables are relevant in the study of non-Gaussianity?
An exactly Gaussian process is one in which all statistical
information is encoded in the two-point expectation value. Any
observation of new information in the three- or higher $n$-point
correlation functions therefore constitutes an observation of
non-Gaussianity.
Other statistical measures exist, such as Minkowski functionals
\cite{Hikage:2008gy}.
However, at the present day the best confrontation of theory with
observation is by means of the
three-point correlation function of the curvature perturbation,
$\zeta$.

To make this quantitative, one defines the bispectrum $B_\zeta$ by
\begin{equation}
    \langle \zeta(\vect{k}_1)
        \zeta(\vect{k}_2)
        \zeta(\vect{k}_3) \rangle
    \equiv (2\pi)^3 \delta(\sum_a \vect{k}_a)
    B_\zeta(\vect{k}_1,\vect{k}_2,\vect{k}_3) ~,
\end{equation}
where $B_\zeta$ is a function of the triangle formed by the three
momenta $\{ \vect{k}_1, \vect{k}_2, \vect{k}_3 \}$, and
we have introduced a notation which will be used throughout
this paper, in which Latin
indices such as $a, b, c, \ldots$ range from $1$ to the number of
momenta carried by the correlation function in question.
These are to be
distinguished from Latin
indices such as $i, j, l, \ldots$, ranging
from $1$ to $3$ which label the spatial dimensions.
It is customary to introduce a dimensionless quantity, $\fnl$,
which generally depends on the momenta $k_a$, and parametrizes the
strength of the non-Gaussian signal. We define
\begin{equation}
    \label{fnl}
    B_\zeta \equiv \frac{6}{5} \fnl
	\sum_{a < b} P_{\zeta}(k_a) P_{\zeta}(k_b)
    \,,
\end{equation}
where $P_{\zeta}(k_1)$ is the power spectrum of $\zeta$, such that
$\langle \zeta(\vect{k}_1) \zeta(\vect{k}_2) \rangle = (2\pi)^3
\delta( \sum_a \vect{k}_a ) P_\zeta(k_1)$. Occasionally, $\fnl$ is
referred to as the \emph{non-linearity parameter}
\cite{Verde:1999ij,Komatsu:2001rj}.
An $\fnl$ that is independent of the momenta corresponds
to a \emph{local} type of non-Gaussianity, which arises from the
real space relation
\begin{equation}
    \zeta = \zeta_g + \frac{3}{5}
    \fnl^\mathrm{local}
    (\zeta_g^2 - \langle \zeta_g^2 \rangle) \;,
    \label{eq:local-bispectrum}
\end{equation}
where $\zeta_g$ is Gaussian. In this case, the bispectrum is maximized
in the limit of one of the three momenta going to zero
\cite{Babich:2004gb}. This form of non-Gaussianity is
produced by nonlinear gravitational evolution subsequent to
horizon crossing
\cite{Babich:2004gb,Lyth:2005fi}.

On the basis of non-linear perturbation theory we might
na\"{\i}vely expect a non-linearity parameter of order unity,
{\ie}, $\fnl \sim \Or(1)$. However (neglecting post-processing in
the later universe from the ubiquitous non-linearities of gravity
\cite{Bartolo:2003gh,Bartolo:2006cu,Bartolo:2006fj,Pitrou:2008ak,
Pitrou:2008hy}) in single-field slow-roll
inflation $\fnl$ is very small---in fact, it is of the same order
as the slow-roll parameters
\cite{Maldacena:2002vr,Acquaviva:2002ud}, which are required to be
much smaller than unity.
Such a small non-linearity
is beyond our present experimental reach, which is
$|\fnl| \gtrsim 100$. However, in the near future the
\emph{Planck} satellite may be sensitive to $\fnl$ of order unity
\cite{Babich:2004yc}.
Any detection of a primordial $|\fnl|$ larger than order unity
will strongly disfavour minimal inflation.

It is possible that interesting information might be encoded in
the higher-order $n$-point functions,
such as the connected part of the four-point function \cite{Hu:2001fa}.
Constraining such an expectation value from CMB data
is extremely challenging and at the time of writing no such
analysis has been completed for the WMAP data, although a study
on small angular scales has been carried out for the BOOMERanG data
\cite{DeTroia:2003tq} and an all-sky 
analysis of the four-point function exists for the older COBE
satellite \cite{Kunz:2001ym}. However, as computing power
increases and algorithms become more efficient, it is quite
plausible that it will be possible to extract four-point
correlations for the WMAP data, and that this will be possible for
\emph{Planck}.
In view of future constraints from CMB experiments and elsewhere
(see for instance
Refs.~\cite{Cooray:2008eb,D'Amico:2007iw}), the purpose of this paper is
to clarify the predictions of single-field slow-roll models of
inflation for the four-point correlation function of the curvature
perturbation $\zeta$.

The four-point correlation function is conventionally parametrized in
terms of the \emph{trispectrum}, written $T_{\zeta}$, and defined by
\cite{Okamoto:2002ik,Kogo:2006kh}
\begin{equation}
    \label{trispectrum_defined}
    \langle \zeta(\vect{k}_1)
        \zeta(\vect{k}_2)
        \zeta(\vect{k}_3)
        \zeta(\vect{k}_4) \rangle
        = (2\pi)^3 \delta(\sum_a \vect{k}_a)
        T_\zeta(\vect{k}_1,\vect{k}_2,\vect{k}_3,\vect{k}_4) \; ,
\end{equation}
where by convention only the connected part of the expectation value on the
left-hand side is considered. The disconnected part is present even in an
exactly Gaussian model, and forms a background from which the genuinely
non-linear signal must be extracted.
Unlike the bispectrum, the trispectrum typically depends on the
relative orientations of the $\vect{k}_a$ in addition to their
magnitudes. If observation forces us to abandon the minimal
scenario of inflation, then the trispectrum is a ready-made
observable which could be consulted for guidance concerning the
modifications which are required.

The trispectrum can be parametrized as \cite{Byrnes:2006vq}
\begin{eqnarray}
    T_\zeta
	\equiv
    \tnl
	\sum_{\substack{b < c \cr a \neq b, c}}
	P_\zeta(k_{ab}) P_\zeta(k_b) P_\zeta(k_c)
	+ \frac{54}{25} \gnl
	\sum_{a < b < c}
	P_\zeta(k_a) P_\zeta(k_b) P_\zeta(k_c)
	\;,
\end{eqnarray}
where $k_{ab} \equiv |\vect k_{ab}|$ and $\vect k_{ab} \equiv
\vect{k}_a + \vect{k}_b $. This definition is motivated by the
fact that, for $\tnl$ and $\gnl$ which are independent of the
momenta, this is the most general \emph{local} parametrization,
corresponding to a non-Gaussianity of the form
\begin{equation}
    \zeta = \zeta_g + \frac{1}2
    \left(\tnl^\mathrm{local}\right)^{1/2}
    \left(\zeta_g^2 - \langle \zeta^2_g \rangle \right) + \frac{9}{25}
    \gnl^\mathrm{local}
    \zeta_g^3 \;
    \label{eq:local-trispectrum}
\end{equation}
in real space. As for the bispectrum, this simple momentum
dependence characterized by constant
$\tnl^\mathrm{local}$ and $\gnl^\mathrm{local}$
is generically produced by gravitational evolution subsequent to
horizon crossing.

More generally, the classical gravitational evolution generates a
curvature perturbation which can be expanded in powers of
$\zeta_g$ with definite
momentum-independent coefficients. In such a case,
if the perturbations are generated by only one field,
one expects that \cite{Byrnes:2006vq}
\begin{equation}
  \tnl^\mathrm{local} = \left(\frac{6}{5}\fnl^\mathrm{local} \right)^2\;.
  \label{local_relation}
\end{equation}
However, the curvature perturbation during inflation is subject to
quantum interference effects around the time of horizon crossing,
and does not admit a simple parameterization in terms of momentum
independent-coefficients \cite{Creminelli:2003iq,Babich:2004gb}.
It follows that in general there
need be \emph{no} relationship between $\tnl$ and $\fnl$.

\begin{figure}
    \vspace{5mm}
    \begin{center}
        \hfill
        \begin{fmfgraph*}(60,60)
            \fmfleft{l}
            \fmfright{r}
            \fmf{fermion,label=$\scriptsize\vect{k}_1$,label.side=left}{l,v1}
            \fmf{fermion,label=$\scriptsize\vect{k}_2$,label.side=left}{v1,r}
            \fmf{fermion,label=$\scriptsize\vect{k}_3$,label.side=left}{r,v2}
            \fmf{fermion,label=$\scriptsize\vect{k}_4$,label.side=left}{v2,l}
            \fmfforce{sw}{l}
            \fmfforce{ne}{r}
            \fmfforce{(0.1w,0.3h)}{v1}
            \fmfforce{(0.75w,0.25h)}{v2}
        \end{fmfgraph*}
        \flabel{a}
        \hfill
        \begin{fmfgraph*}(60,60)
            \fmfleft{l}
            \fmfright{r}
            \fmf{fermion,label=$\scriptsize\vect{k}_1$,label.side=left}{l,v1}
            \fmf{fermion,label=$\scriptsize\vect{k}_2$,label.side=left}{v1,r}
            \fmf{fermion,label=$\scriptsize\vect{k}_3$,label.side=left}{r,v2}
            \fmf{plain,label=$\scriptsize\vect{k}_4$,label.side=left}{v2,l}
            \fmfforce{sw}{l}
            \fmfforce{ne}{r}
            \fmfforce{(0.2w,0.7h)}{v1}
            \fmfforce{(0.05w,0h)}{v2}
        \end{fmfgraph*}
        \flabel{b}
        \hfill
        \begin{fmfgraph*}(60,60)
            \fmfleft{l}
            \fmfright{r}
            \fmf{plain,label=$\scriptsize\vect{k}_1$,label.side=left}{l,v1}
            \fmf{plain,label=$\scriptsize\vect{k}_3$,label.side=left}{v1,r}
            \fmf{plain,label=$\scriptsize\vect{k}_4$,label.side=left}{r,v2}
            \fmf{plain,label=$\scriptsize\vect{k}_2$,label.side=left}{v2,l}
            \fmfforce{sw}{l}
            \fmfforce{ne}{r}
            \fmfforce{(0.3w,0.6h)}{v1}
            \fmfforce{(0.35w,0.55h)}{v2}
        \end{fmfgraph*}
        \flabel{c}
        \hfill
        \begin{fmfgraph*}(60,60)
            \fmfleft{l}
            \fmfright{r}
            \fmf{fermion,label=$\scriptsize\vect{k}_1$,label.side=left}{l,v1}
            \fmf{fermion,label=$\scriptsize\vect{k}_3$,label.side=left}{v1,r}
            \fmf{fermion,label=$\scriptsize\vect{k}_2$,label.side=left}{r,v2}
            \fmf{fermion,label=$\scriptsize\vect{k}_4$,label.side=left}{v2,l}
            \fmfforce{sw}{l}
            \fmfforce{ne}{r}
            \fmfforce{(0.3w,0.6h)}{v1}
            \fmfforce{(0.7w,0.4h)}{v2}
        \end{fmfgraph*}
        \flabel{d}
        \hfill
        \mbox{}
    \end{center}
    \caption{\label{fig:kites}
    Possible planar momentum quadrilaterals.
    Note, however, that there is no need for the quadrilateral to lie in
	a plane,
    and in general it is a three-dimensional object.
    In \flabel{a}, there is no particular relationship among the sides of
    the quadrilateral. This is the general case.
    In \flabel{b}, the side associated with $\vect{k}_4$ is taken to
    zero length, causing the quadrilateral to degenerate to a triangle.
    In \flabel{c}, adjacent sides possess equal magnitudes and their
    directions are becoming opposite, a limit we refer to as the
    \emph{folded kite}.
    In \flabel{d}, opposite sides possess equal magnitudes,
    yielding a parallelogram. The shapes described in \flabel{c} and
	\flabel{d}, are therefore precisely dual to each other,
    and describe the same physics.
    }
\end{figure}
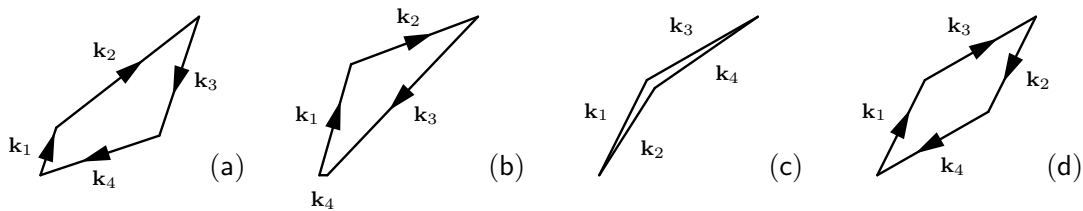

There are two interesting degenerate configurations where a
non-Gaussianity of local form maximizes the trispectrum.
The first is the so-called ``squeezed''
limit, where one of the four momenta is taken to zero.
In this limit, the quadrilateral formed by the momentum vectors
$\vect{k}_a$ degenerates into a triangle.
(See Fig.~\ref{fig:kites}\flabel{a}--\flabel{b}.)
For instance, by taking $k_4 \ll k_1, k_2, k_3$ one obtains
\begin{equation}
    \fl
    T^\mathrm{local}_\zeta
	=
    \left( 2 \tnl^\mathrm{local} + \frac{54}{25} \gnl^\mathrm{local} \right)
	P_\zeta(k_4) \left[
    P_\zeta(k_1) P_\zeta(k_2) + P_\zeta(k_1) P_\zeta(k_3) +
	P_\zeta(k_2) P_\zeta(k_3) \right]\;.
\end{equation}
Thus,
$\tnl^\mathrm{local}$ and $\gnl^\mathrm{local}$
contribute equally to the trispectrum in this
limit.

The second interesting limit occurs when the magnitude of the sum
of two momenta is taken to zero, so that $k_{ij} = | \vect{k}_i +
\vect{k}_j | \rightarrow 0$ for some $i \neq j$. We refer to this
limit as the \emph{counter-collinear} case
because for each momentum there is another one with equal
magnitude and opposite direction.
One can visualize the quadrilateral formed by the momentum vectors
in this limit in two different but equivalent ways. If we choose
to order the momenta so that the counter-collinear pairs are
adjacent, then we generate what can be called a ``folded kite,''
shown in Fig.~\ref{fig:kites}\flabel{c}. On the other hand, if we
choose to order the momenta so that the counter-collinear pairs are
on opposite sides of the quadrilateral, then it becomes
\emph{planar} and yields a parallelogram, as in
Fig.~\ref{fig:kites}\flabel{d}. Changing the order of the momenta
does not affect the trispectrum, which is a function only of the
intrinsic geometry of the $\vect{k}_a$; for the trispectrum, this
means it depends on a total of three magnitudes and three angles.
Therefore, the planar shapes given in
Fig.~\ref{fig:kites}\flabel{c}--\flabel{d} are precisely dual to
each other.
The local trispectrum simplifies in the counter-collinear limit. By
taking $k_{12} \ll k_1 \approx k_2, k_3 \approx k_4$
one obtains
\begin{equation}
    \label{local_2}
    T^\mathrm{local}_\zeta (\vect{k}_1,\vect{k}_2,\vect{k}_3,\vect{k}_4) =
    4 \tnl^\mathrm{local} P_\zeta(k_{12})
    P_\zeta(k_1) P_\zeta(k_3) \;.
\end{equation}
Thus, only $\tnl^\mathrm{local}$
contributes to the trispectrum in this limit.

As discussed above,
it does not always happen that a model of the early universe
predicts a curvature perturbation whose non-linearities are
precisely of the local form described by
Eqs.~\eref{eq:local-bispectrum} and~\eref{eq:local-trispectrum}.
To accommodate this situation, one can take $\tnl$ and $\gnl$ as
momentum dependent; however, in this case the separation between
$\tnl$ and $\gnl$ becomes ambiguous. Indeed, several authors
\cite{Boubekeur:2005fj,Lyth:2005fi,Alabidi:2005qi} have used a
momentum dependent $\tnl$ alone to parametrize the full
trispectrum. The division of the trispectrum into terms
proportional to $\tnl$ and $\gnl$ is helpful where these
parameters can be made approximately momentum-independent, but
this is not always the case when the non-Gaussianity is dominated
by quantum interference effects around the time of horizon exit,
as in single-field inflation. For this reason we will mostly avoid
the use of $\tnl$ or $\gnl$ in this paper, giving expressions for
the trispectrum directly. However, as we will see, the
contribution of graviton exchange to the trispectrum is dominant
in the parallelogram limit, in such a way that we can describe it
as a momentum independent $\tnl^\mathrm{local}$.

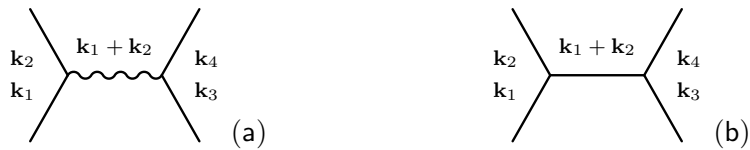
\begin{figure}
    \begin{center}
        \hfill
        \begin{fmfgraph*}(80,50)
            \fmfpen{thin}
            \fmfleft{l1,l2}
            \fmfright{r1,r2}
            \fmf{plain,label=$\scriptsize\vect{k}_1$,label.side=left}{l1,v1}
            \fmf{plain,label=$\scriptsize\vect{k}_2$,label.side=right}{l2,v1}
            \fmf{wiggly,label=$\scriptsize\vect{k}_1+\vect{k}_2$,
                label.dist=0.15h,tension=0.8}{v1,v2}
            \fmf{plain,label=$\scriptsize\vect{k}_3$}{r1,v2}
            \fmf{plain,label=$\scriptsize\vect{k}_4$}{r2,v2}
        \end{fmfgraph*}
        \flabel{a}
        \hfill
        \begin{fmfgraph*}(80,50)
            \fmfpen{thin}
            \fmfleft{l1,l2}
            \fmfright{r1,r2}
            \fmf{plain,label=$\scriptsize\vect{k}_1$,label.side=left}{l1,v1}
            \fmf{plain,label=$\scriptsize\vect{k}_2$,label.side=right}{l2,v1}
            \fmf{plain,label=$\scriptsize\vect{k}_1+\vect{k}_2$,
                 label.dist=0.15h,tension=0.8}{v1,v2}
            \fmf{plain,label=$\scriptsize\vect{k}_3$}{r1,v2}
            \fmf{plain,label=$\scriptsize\vect{k}_4$}{r2,v2}
        \end{fmfgraph*}
        \flabel{b}
        \hfill
        \mbox{}
    \end{center}
    \caption{\label{fig:graviton-exchange}In \flabel{a},
    exchange of a graviton (represented by a wavy line)
    leads to correlations among four scalar
    fluctuations (represented by straight lines).
    In the limit where the graviton which is exchanged becomes
    extremely soft, so that $|\vect{k}_1 + \vect{k}_2| \rightarrow 0$,
    one can think of such correlations
    as being mediated
    by fluctuations on top of a modified background which carries
    a classical gravitational wave. The same interpretation holds in
    \flabel{b}, where the exchange is mediated by a scalar particle.
    }
\end{figure}

In perturbation theory our expectation would again be $\tnl \sim \gnl
= \Or(1)$. However, it has been shown that, in slow-roll inflation
with canonical kinetic term,%
    \footnote{The calculation of the trispectrum in single-field models with
    non-canonical kinetic terms was first performed,
	neglecting metric perturbations, in
    Ref.~\cite{Huang:2006eh} and later, more generally, in
    Ref.~\cite{Arroja:2008ga}. In these models $\tnl$ can be of order unity
    or larger. The contribution from the graviton exchange is suppressed
    by slow-roll with respect to the one due to the contact interaction
    and it has been neglected in those references. Note that the
    calculation performed in this paper, {\ie}, of the trispectrum from
    slow-roll inflation models based on a field with standard kinetic term,
    requires one to pursue the calculation to one order higher in slow-roll
    than was done in Refs. \cite{Huang:2006eh,Arroja:2008ga}.
    Thus, those references do not include our final results.}
$\tnl$ exhibits a similar suppression by one power of
the slow-roll parameter $\epsilon \equiv -\dot H/H^2$, giving
$\tnl \lesssim r/50$ \cite{Seery:2006vu}, where $r$ is the ratio of
tensor to scalar amplitudes, $r \equiv 8 P_T/P_{\zeta}$,
and in single-field inflation $r = 16 \epsilon$.
It follows that the non-linearity parameter $\tnl$ is roughly of
the same size as $\fnl$. (In practice, of course, the trispectrum
is much harder to detect than the bispectrum because an extra
power of the power spectrum $P_\zeta$ is equivalent to suppression
by $\sim 10^{-10}$.) However, one may have some reservations about
this conclusion.  In particular, the calculation reported in
Ref.~\cite{Seery:2006vu} ignored nonlinear interactions mediated
by scalar or tensor fluctuations.  In terms of diagrams, this is
equivalent to neglecting the two diagrams where the correlation
among the four scalar modes is due to the exchange of scalar
or tensor quanta. The latter diagram, which
turns out to be more important, is shown in
Fig.~\eref{fig:graviton-exchange}\flabel{a}, whereas the scalar
exchange is described in
Fig.~\ref{fig:graviton-exchange}\flabel{b}.

The vertices of these nonlinear interaction can be computed
straightforwardly using the action for Einstein gravity coupled to
a single scalar field, which was studied to third order
by Maldacena \cite{Maldacena:2002vr}. For the vertex which describes
a contact interaction
between three quanta of the curvature perturbation, $\zeta$,
the appropriate term in the Lagrangian is typically
suppressed by at least one power of the slow-roll parameters,
{\ie},~${\cal L}_{\zeta \zeta \zeta} / {\cal L}_{\zeta \zeta}
\sim \Or(\epsilon,\eta) \cdot P_{\zeta}^{1/2}$ and
therefore
\begin{center}
    \parbox{15mm}{\begin{fmfgraph*}(40,30)
        \fmfpen{thin}
        \fmfleft{l}
        \fmfright{r1,r2}
        \fmf{plain}{l,v}
        \fmf{plain}{r1,v}
        \fmf{plain}{r2,v}
    \end{fmfgraph*}}
    $\sim \Or (\epsilon,\eta)$ ,
\end{center}
where $\eta$ is the second slow-roll parameter, defined as $\eta
\equiv \Planck^2 V''/V$,
where we have introduced the reduced Planck mass $\Planck \equiv (8
\pi G)^{-1/2}$.
Thus, bearing in mind that a four-point correlation established by
scalar exchange must contain two vertices of this sort, we expect
that the contribution to the trispectrum from this process is of
the order of \emph{two} powers of slow-roll, {\ie}, $\tnl \sim
\Or(\epsilon^2,\eta^2, \epsilon \eta)$. This is suppressed by one
power of slow-roll compared with the scalar contact interaction
and can therefore be ignored. However, things are different when
we consider the exchange of a graviton. In this case the
interaction Lagrangian between a graviton and two scalars is not
suppressed by any powers of slow-roll parameters at all,
{\ie},~${\cal L}_{\gamma \zeta \zeta} / {\cal L}_{\zeta \zeta}
\sim P_{\gamma}^{1/2}$ and therefore
\begin{center}
    \parbox{15mm}{\begin{fmfgraph*}(40,30)
        \fmfpen{thin}
        \fmfleft{l}
        \fmfright{r1,r2}
        \fmf{wiggly}{l,v}
        \fmf{plain}{r1,v}
        \fmf{plain}{r2,v}
    \end{fmfgraph*}}
    $\sim r^{1/2}$ .
\end{center}
The power spectrum of gravitational waves is smaller than that of
$\zeta$ by a factor of $r$,
and therefore this contribution is of order $r$.
It follows that the effect of graviton exchange is comparable to
the contact contribution studied in Ref.~\cite{Seery:2006vu}.
This observation was first made by Arroja \& Koyama \cite{Arroja:2008ga}.%
    \footnote{In Ref.~\cite{Arroja:2008ga} it was
    also claimed that the discussion of Ref.~\cite{Jarnhus:2007ia} neglected
    tensor modes. However, the tensor contribution originally derived by
    Maldacena \cite{Maldacena:2002vr}, which was discussed again in
    Ref.~\cite{Arroja:2008ga}, was in fact included in
    Ref.~\cite{Jarnhus:2007ia}, where it was shown to confirm their
    conclusions.

    One might also worry that the graviton--scalar--scalar
    vertex could dominate corrections to the scalar power spectrum, which
    arise via loops which exchange virtual quanta.
    The correction to the scalar
    power spectrum due to loops from the graviton--scalar--scalar
    interaction has been computed in \cite{Dimastrogiovanni:2008af},
    where it was shown that it is of the same magnitude as that coming
    from scalar self-interactions
    \cite{Sloth:2006az,Sloth:2006nu,Seery:2007we,Seery:2007wf,
    Urakawa:2008rb}.}

In the present paper, we will calculate the contribution to
the four-point function of scalar curvature perturbations which arises
from correlations mediated by graviton exchange. We explicitly carry out this
calculation for single-field models, but our result is easily generalized
to many fields and our final results will apply to multi-field examples.
Like many previous
calculations of the non-linearities among scalar fluctuations at
horizon crossing, this calculation is somewhat involved.
Fortunately, in the counter-collinear
limit ({\ie}, parallelogram or folded-kite), the
amplitude and momentum dependence of the contribution to the
trispectrum from graviton exchange can be computed
straightforwardly by a simple semiclassical argument, which provides
an important check of our computation.

In the next section we explicitly compute the contribution to the
trispectrum from graviton exchange, using the uniform curvature gauge,
while in \S\ref{sec:consistency} we discuss the counter-collinear limit
and check our calculation.  In \S\ref{sec:tot}
we give an expression for the total trispectrum, including the
contribution of Ref.~\cite{Seery:2006vu} and in \S\ref{sec:conclude}
we conclude with a discussion.
Finally, in~\ref{appendix:integral} we report the calculation of a
long double integral used in \S\ref{sec:trispectrum}.

Throughout this paper, we use fundamental units in which Planck's constant
$\hbar$ and the speed of light $c$ are set to unity, so $\hbar = c = 1$.
We choose to set
$\Planck^{-2} \equiv 8 \pi G$ equal to unity. Our metric
convention is $(-,+,+,+)$.

%%%%%%%%%%%%%%%%%%%%%%%%%%%%%%%%%%%%%%%%%%%%%%%%%%%%%%%%%%%%%%%%%%%%%%%%%%%%
%%%%%%%%%%%%%%%%%%%%%%%%%%%%%%%%%%%%%%%%%%%%%%%%%%%%%%%%%%%%%%%%%%%%%%%%%%%%

\section{Trispectrum from graviton exchange}
\label{sec:trispectrum}

%%%%%%%%%%%%%%%%%%%%%%%%%%%%%%%%%%%%%%%%%%%%%%%%%%%%%%%%%%%%%%%%%%%%%%%%%%%%
%%%%%%%%%%%%%%%%%%%%%%%%%%%%%%%%%%%%%%%%%%%%%%%%%%%%%%%%%%%%%%%%%%%%%%%%%%%%

In this section we want to evaluate the contribution to the
four-point function of curvature perturbations from the exchange
of a graviton. This is the process shown in
Fig.~\ref{fig:graviton-exchange}\flabel{a}, which involves a
third-order interaction among scalar fluctuations and tensor
perturbations of the metric. The description of this interaction
requires that we expand the action to third order in such
fluctuations. For simplicity, we will restrict to a single scalar
field only. The calculation can easily be extended to more than
one field, and indeed our final result for the trispectrum of the
curvature perturbation $\zeta$ remains valid in the multi-field case
with flat field space metric.

We begin with Einstein gravity coupled to a scalar field, which gives a
total action
\begin{equation}
    S = \frac{1}{2}\int \sqrt{-g}\left[R -(\nabla
    \phi)^2-2V(\phi)\right]~,
    \label{eq:action}
\end{equation}
where $R$ is the spacetime Ricci scalar and $V(\phi)$ is an arbitrary
potential.
Using the Arnowitt--Deser--Misner method (or so-called ``ADM formalism'')
\cite{Arnowitt:1960es},
one can derive the action for coupled scalar and perturbations
from Eq.~\eref{eq:action} to any given order using an iterative procedure.
The ADM line element is given by
\begin{equation}
    ds^2= -N^2 dt^2 + a^2(t) h_{ij}(dx^i + N^idt)(dx^j + N^jdt)~,
\end{equation}
where $a(t)$ is the scale factor, and $N$, $N^i$ are the lapse and
the shift functions. The three-dimensional metric $h_{ij}$ encodes
scalar and tensor fluctuations in the spatial geometry. When the
ADM metric ansatz is inserted into Eq.~\eref{eq:action}, the lapse
and the shift functions act as Lagrange multipliers: the field
equations obtained by minimizing the action in $N$ and $N^i$ give
the constraint equations which follow from the Einstein equation,
while $\phi$ and $h_{ij}$ are the dynamical degrees of freedom.

As in Ref. \cite{Seery:2006vu}, we choose to work in the uniform
curvature gauge. In this gauge we take the scalar field and the
spatial metric to be parametrized by
\begin{equation}
    \phi = \phi(t) + \dpp (t, \vect x)~, \qquad
    h_{ij}=a(t)(e^{\gamma})_{ij}~.
\end{equation}
Here $\dpp(t, \vect{x})$ is the inflaton field fluctuation, and
$\gamma_{ij}$ is a traceless and divergenceless tensor
fluctuation, obeying $\partial_i\gamma_{ij}=\gamma_{ii}=0$. Our
eventual aim is to compute the four-point function of the comoving
curvature perturbation, $\R$, which on super-Hubble scales is
equal to the curvature perturbation on uniform density slices,
$\zeta$ \cite{Lyth:2004gb,Vernizzi:2004nc,Langlois:2006vv}.
However, the calculation is considerably simplified if we choose
to calculate the four-point function of $\dpp$ at horizon crossing
and then relate it to the four-point function of the uniform
density curvature perturbation $\zeta$.
Indeed, as shown in Ref.~\cite{Maldacena:2002vr},
in virtue of the smallness of the slow-roll parameters at
Hubble crossing, it is possible to reproduce the correct
late-time behaviour of $\zeta$ using the free-field
solution in de Sitter spacetime for $\varphi$.
At late times $\zeta$ is constant and, at lowest order
in slow-roll, it can simply be normalized to $\varphi$
at horizon crossing.

The free action, which describes the evolution of independent
scalar and tensor fluctuations, is
\begin{equation}
    \fl
    S_2 = \int d^3 x \, dt \; a^3 \left[ \frac{1}{2} \left( \dot{\dpp }^2 -
    \frac{\partial_i \dpp \partial_i \dpp}{a^2} \right) + \frac{1}{8}
    \left( \dot{\gamma}_{ij} \dot{\gamma}_{ij} - \frac{\partial_k
    \gamma_{ij} \partial_k \gamma_{ij}}{a^2} \right) \right]\; ,
\end{equation}
where we are using the convention that a repeated spatial index
in the lowered position, such
as $i$ or $j$, denotes summation with the flat
spatial metric $\delta_{ij}$.
The free fields derived from this action satisfy the equations of motion
obtained by varying $\dpp$ and $\gamma_{ij}$, and can therefore be built
out of mode functions $U_k(t)$, $U^\ast_k(t)$ and
$\gamma_k^s(t)$, $\gamma_k^{s \ast}(t)$ which solve the classical
equations of motion,
\begin{eqnarray}
    U_k(\eta) & = \frac{H_*}{\sqrt{2k^3}}(1-ik\eta)e^{ik\eta}\;,
    \label{free_p} \\
    \gamma^s_k (\eta) & =
    \frac{H_*}{\sqrt{k^3}}(1-ik\eta)e^{ik\eta}
    \quad \mbox{(any $s$)} \;, \label{free_g}
\end{eqnarray}
where $\eta$ is the conformal time, defined by $d \eta =d t/a$,
and $H_*$ is the Hubble parameter evaluated at Hubble crossing for
the mode $k$; an  asterisk ``$\ast$'' as superscript denotes
complex conjugation. The index $s$ labels the two possible
polarization states ``$+$'' and ``$\times$'' of a spin two
gravitational excitation. Introducing creation and annihilation
operators $a^+_{\vect{k}}, a^-_{\vect{k}}$ for the scalar quanta
we obtain
\begin{equation}
    \dpp (x)    = \int \frac{d^3 k}{(2\pi)^{3}}
    \left[a^-_{\vect{k}}U_k(t)+a^+_{-\vect{k}}U^*_k(t)\right]
    e^{i\vect{k}\cdot\vect{x}}\;,
\end{equation}
whereas, with corresponding operators
$b^{+s}_{\vect{k}}$, $b^{-s}_{\vect{k}}$ for each polarization of the
gravitational fluctuation, we find
\begin{equation}
    \gamma_{ij}(x) = \sum_{s=+,\times} \int
    \frac{d^3 k}{(2\pi)^{3}}\left[
        b^{-s}_{\vect{k}}\ep^s_{ij}(\vect{k})\gamma^s_k(t)
        + b^{+s}_{-\vect{k}}\ep_{ij}^s(-\vect{k})\gamma^{s*}_k(t)\right]
        e^{i\vect{k}\cdot\vect{x}}~.
\end{equation}
The polarization tensors
$\ep^s_{ij}$ are chosen to satisfy
the transversality and tracelessness conditions
$\ep^s_{ii}(\vect k)=k^i\ep^s_{ij}(\vect k)=0$,
together with a completeness relation obtained by tracing over
spatial indices, $\ep_{ij}^s(\vect{k})
\ep_{ij}^{\ast s'}(\vect{k})=2\delta_{ss'}$

We have now reduced the problem to calculating the four-point
function of fluctuations $\dpp$ in the inflaton field at horizon
crossing, where the correlations are produced via the graviton
exchange diagram. To proceed, we need only the relevant third
order term in the action, which describes the
graviton--scalar--scalar vertex in uniform curvature gauge.
This is \cite{Maldacena:2002vr},
\begin{equation}
    S_3 = \frac{1}{2}\int d^3x \, d\eta \; a^2 \gamma^{ij} \partial_i \dpp
    \partial_j \dpp\;.
    \label{eq:interaction}
\end{equation}

The correct approach to computing expectation values in an expanding
or curved background spacetime is the so-called \emph{in--in} or
Schwinger--Keldysh formalism
\cite{Schwinger:1960qe,Keldysh:1964ud,Calzetta:1986ey}.
In this formalism the expectation value of any operator $\mathcal{O}$
(evaluated at time $\eta_\ast$) is given by
\begin{equation}
    \fl\label{exp1}
    \lle \Omega \rb \mathcal{O}(\eta_*) \lb \Omega \rgr = \frac{\lle
    0\rb
    \Torder \left\{\mathcal{O}(\eta_*)~e^{-i\int_{-\infty}^{\eta_*}d\eta\left[
    \Hint(\phi,\dpp ^+,\gamma^+)- \Hint(\phi,\dpp
    ^-,\gamma^-)\right]}\right\}\lb 0\rgr}{\lle 0\rb
    \Torder \left\{e^{-i\int_{-\infty}^{\eta_*}d\eta\left[ \Hint(\phi,\dpp
    ^+,\gamma^+)- \Hint(\phi,\dpp ^-,\gamma^-)\right]}\right\}\lb
    0\rgr}~,
\end{equation}
where $\lb \Omega \rgr$ is the vacuum of the interacting theory
while $\lb 0\rgr$ is the
vacuum of the free theory,
$\Torder$ is the time ordering operator and $\Hint$ is the
interaction Hamiltonian given, in our case, by%
    \footnote{In general, the interaction Hamiltonian does not have
    to be given by the sign reverse of the interaction Lagrangian.
    This is only true in a limited class of theories without derivative
    interactions, which does not include the case of
    coupled matter and metric
    fluctuations during inflation. It follows that extra terms can be
    generated in the transformation to the interaction Hamiltonian
	\cite{Weinberg:2005vy,Adshead:2008gk}.
    For models with non-minimal kinetic terms, this can lead to
    non-negligible contributions to the trispectrum, as discussed
    in \texttt{v3} of Ref.~\cite{Huang:2006eh}. In the case of
    slow-roll inflation with canonical kinetic terms, however, these
    effects can be ignored.
	}
\begin{equation}
    \Hint = - \frac{1}{2}\int d^3 x \; a^2 \gamma^{ij} \partial_i \dpp
    \partial_j \dpp\;.
    \label{eq:hint}
\end{equation}
The matrix element~\eref{exp1} describes a system
evolved from its initial vacuum state at $\eta \to -\infty$ to $\eta =\eta_*$
with an
operator inserted at $\eta$, and back again from $\eta_*$ to $-\infty$,
with a set of ``$+$'' fields on the increasing-time contour and a
set of ``$-$'' fields on the decreasing-time contour. Eventually, $\eta_*$
will be taken to $0$ and
the final result is finite in this limit.

The contractions between different pairs of the two types of
field perturbations yield four kinds of propagator,
\begin{equation}
    G^{\pm\pm}(x,x') \equiv \lle 0\rb \Torder \left[\dpp ^{\pm}(x)\dpp
    ^{\pm}(x')\right]\lb 0\rgr\;.\label{TO_dp}
\end{equation}
It is convenient to rewrite this expression in Fourier space, where the
propagator takes the form
\begin{equation}
    G^{\pm\pm}(x,x')=
    \int\frac{d^3k}{(2\pi)^3}e^{i\vect{k}\cdot(\vect{x}-\vect{x}')}
    G_k^{\pm\pm}(\eta,\eta')~,\label{TO_dp_F}
\end{equation}
and $G_k^{\pm\pm}$ can be written as
\begin{eqnarray}
    G^{++}_k(\eta,\eta') & =
        G^{>}_k(\eta,\eta')\Theta(\eta-\eta') +
        G^{<}_k(\eta,\eta')\Theta(\eta'-\eta) \; , \nonumber \\
    G^{--}_k(\eta,\eta') & =
        G^{>}_k(\eta,\eta')\Theta(\eta'-\eta) +
        G^{<}_k(\eta,\eta')\Theta(\eta-\eta') \; , \nonumber \\
    G^{-+}_k(\eta,\eta') & =
        G^{>}_k(\eta,\eta') \; , \nonumber \\
    G^{+-}_k(\eta,\eta') & =
        G^{<}_k(\eta,\eta') ~ .
\end{eqnarray}
The ``$>$'' and ``$<$'' functions are defined by
\begin{eqnarray}
    G^{>}_k(\eta,\eta') & = U_k(\eta)U^*_k(\eta') \; ,\nonumber\\
    G^{<}_k(\eta,\eta') & = U^*_k(\eta)U_k(\eta') ~ . \label{G_p}
\end{eqnarray}
Similarly, for the tensor modes we have a propagator
$F_{ijlm}^{ss' \pm\pm}$
\begin{equation}
    F_{ijlm}^{ss'\pm\pm}(x,x') \equiv \lle 0\rb
        \Torder [
            \gamma_{ij}^{s\pm}(x)\gamma_{lm}^{s'\pm}(x')
        ] \lb 0 \rgr ~,
    \label{TO_gamma}
\end{equation}
which goes over to the Fourier space representation
\begin{eqnarray}
    F_{ijlm}^{ss'\pm\pm}(x,x') =
    \sum_{s = +,\times}
    \int\frac{d^3k}{(2\pi)^3}e^{i\vect{k}\cdot(\vect{x}-\vect{x}')}
    \ep^s_{ij}(\vect k) \ep^s_{lm}(-\vect k)
    F_k^{s\pm\pm}(\eta,\eta')~.\label{TO_gamma_F}
\end{eqnarray}
As before, explicit expressions can be given in terms of
``$>$'' and ``$<$'' functions, leading to
\begin{eqnarray}
    F^{s++}_k(\eta,\eta') & =
        F^{s>}_k(\eta,\eta')\Theta(\eta-\eta') +
        F^{s<}_k(\eta,\eta')\Theta(\eta'-\eta) \; , \nonumber \\
    F^{s--}_k(\eta,\eta') & =
        F^{s>}_k(\eta,\eta')\Theta(\eta'-\eta) +
        F^{s<}_k(\eta,\eta')\Theta(\eta-\eta') \; , \nonumber \\
    F^{s-+}_k(\eta,\eta') & =
        F^{s>}_k(\eta,\eta') \; , \nonumber \\
    F^{s+-}_k(\eta,\eta') & =
        F^{s<}_k(\eta,\eta')~,
\end{eqnarray}
where one has
\begin{eqnarray}
    F^{s>}_k(\eta,\eta') & = \gamma^s_k(\eta)\gamma^{s*}_k(\eta')
        \; , \nonumber \\
    F^{s<}_k(\eta,\eta') & = \gamma^{s*}_k(\eta)\gamma^{s}_k(\eta')~.
    \label{F_g}
\end{eqnarray}

Expanding the exponential in Eq.~\eref{exp1} to second order,
using the interaction Hamiltonian given by Eq.~\eref{eq:hint}, and
making all possible Wick contractions among the fields gives the
contribution to the four-point function from graviton exchange
(which we label with the superscript ``GE''). Since there are two
vertices in the diagram, the result depends on two time integrals.
Using Eqs.~\eref{TO_dp} and~\eref{TO_gamma} to carry out the Wick
contractions, together with the Fourier space
definitions~\eref{TO_dp_F} and~\eref{TO_gamma_F}, leads to
\begin{eqnarray}
    \fl\nonumber
    \langle
        \dpp_{\bk_1}\dpp_{\bk_2}\dpp_{\bk_3}\dpp_{\bk_4}
    \rangle_*^\mathrm{GE}
    =
    - \frac{1}{4} (2\pi)^3\delta^3 (\sum_a\vect{k}_a)
    \sum_s \ep_{ij}^s(\vect k_{12}) k_1^i k_2^j
           \ep_{lm}^s(\vect k_{34}) k_3^l k_4^m
    \int_{-\infty}^{\eta_*} \frac{d\eta}{\eta^2}
    \int_{-\infty}^{\eta}   \frac{d\eta'}{\eta'^2} \\ \nonumber
    \mbox{} \times \Big[
        G^{++}_{k_1}(\eta_*,\eta) G^{++}_{k_2}(\eta_*,\eta)
        G^{++}_{k_3}(\eta_*,\eta') G^{++}_{k_4}(\eta_*,\eta')
        F^{s++}_{k_{12}}(\eta,\eta') \\ \nonumber
        \hspace{2mm} \mbox{} +
        G^{+-}_{k_1}(\eta_*,\eta) G^{+-}_{k_2}(\eta_*,\eta)
        G^{+-}_{k_3}(\eta_*,\eta') G^{+-}_{k_4}(\eta_*,\eta')
        F^{s--}_{k_{12}}(\eta,\eta') \\ \nonumber
        \hspace{2mm} \mbox{} -
        G^{++}_{k_1}(\eta_*,\eta) G^{++}_{k_2}(\eta_*,\eta)
        G^{+-}_{k_3}(\eta_*,\eta') G^{+-}_{k_4}(\eta_*,\eta')
        F^{s+-}_{k_{12}}(\eta,\eta') \\ \nonumber
        \hspace{2mm} \mbox{} -
        G^{+-}_{k_1}(\eta_*,\eta) G^{+-}_{k_2}(\eta_*,\eta)
        G^{++}_{k_3}(\eta_*,\eta') G^{++}_{k_4}(\eta_*,\eta')
        F^{s-+}_{k_{12}}(\eta,\eta')
    \Big] \\
    \mbox{} + \mbox{23 permutations}\;.
\end{eqnarray}
Using
$G_k^{++}(\eta_*,\eta) = G^>(\eta_*,\eta)$ for
$\eta_*>\eta$, $F^{++}_k(\eta,\eta')=F^{>}_k(\eta,\eta')$ for
$\eta>\eta'$, $G^> = G^{<*}$ and other equivalent relations, this can
be simplified to give
\begin{eqnarray}
    \fl \left< \dpp_{{\vect k}_1}\dpp_{{\vect k}_2}
    \dpp_{{\vect k}_3}\dpp_{{\vect k}_4}\right>_*^\mathrm{GE} =
    (2\pi)^3\delta^3(\sum_a \vect{k}_a)\sum_s \ep_{ij}^s(\vect
    k_{12})k_1^ik_2^j\ep_{lm}^s(\vect k_{34})k_3^l k_4^m
    \int_{-\infty}^{\eta_*}\frac{d\eta}{\eta^2}\int_{-\infty}^{\eta}
    \frac{d\eta'}{\eta'^2}\nonumber\\
    \mbox{} \times
    \Im\left[G^{>}_{k_1}(\eta_*,\eta)G^{>}_{k_2}(\eta_*,\eta)\right]
    \cdot
    \Im\left[G^{>}_{k_3}(\eta_*,\eta')G^{>}_{k_4}(\eta_*,\eta')
    F^{s>}_{k_{12}}(\eta,\eta')
    \right] \nonumber\\
    \mbox{} + \mbox{23 permutations}\;.
    \label{ge_before_simplifications}
\end{eqnarray}

To proceed, one makes use of Eqs.~\eref{G_p} and~\eref{F_g} for
the propagators $G_k^{>}(\eta_*,\eta')$ and $F_k^{s>}(\eta,\eta')$
in this equation, together with the free mode functions for the
scalar field and tensor fluctuations, Eqs.~\eref{free_p} and
\eref{free_g}. In total, for the four-point function we obtain,%
    \footnote{An equivalent---and computationally equivalent---%
	interaction picture formalism was
    derived by Weinberg \cite{Weinberg:2005vy}. With this formalism
    the trispectrum can be computed straightforwardly
	using Eq.~(1) of this Ref., expanded to second order
	in the interactions.}
\begin{eqnarray}
    \fl\nonumber \langle \dpp _{\bk_1}\dpp _{\bk_2}\dpp _{\bk_3}\dpp_{\bk_4}
    \rangle_*^\mathrm{GE}
    = (2\pi)^3 \delta (\sum_a \vect k_a)
    \frac{ H_*^6}{\prod_a(2 k_a^3) k_{12}^3} \cdot \sum_s
    \epsilon_{ij}^s(\vect{ k}_{12}) \epsilon_{lm}^s(\vect k_{34})
    k_1^i k_2^j k_3^l k_4^m \cdot {\cal I}_{1234}
    \\ \mbox{} + \mbox{23 permutations} \;, \label{4p-1}
\end{eqnarray}
where ${\cal I}_{1234}$ is defined as
\begin{eqnarray}
    \fl\nonumber
    \label{int_1234}
    {\cal I}_{1234} \equiv
    \int^{\eta_*}_{-\infty} \frac{d\eta}{\eta^2} \int^\eta_{-\infty}
    \frac{d\eta'}{\eta'^2}
    \Im
        \Big  [ (1+ik_1 \eta_*) (1+ik_2 \eta_*)
                (1-ik_1 \eta) (1-ik_2 \eta) e^{i(\eta-\eta_\ast)(k_1+k_2)}
        \Big  ]
    \\ \nonumber
    \mbox{} \times
    \Im
        \Big  [ (1+ik_3 \eta_*) (1+ik_4 \eta_*)
                (1-ik_3 \eta')(1-ik_4 \eta')
                \\ \qquad\quad \mbox{} \times
                (1+ik_{12} \eta)
                (1-ik_{12} \eta') e^{i(\eta'-\eta_\ast)(k_3+k_4)}
                e^{-i(\eta-\eta')k_{12}}
        \Big  ] \;,
\end{eqnarray}
and $k_t$ is the sum of the magnitudes of the momenta, $k_t \equiv \sum_a k_a$.

Let us work out the permutations in more detail. Since
$\epsilon^s_{ij}$ is a symmetric tensor and Eq.~\eref{4p-1} is
invariant under the exchanges $1 \leftrightarrow 2$ and $3
\leftrightarrow 4$, one can rewrite this equation as
\begin{eqnarray}
    \fl\nonumber
    \langle
        \dpp _{\bk_1}\dpp _{\bk_2}\dpp _{\bk_3}\dpp_{\bk_4}
    \rangle_*^\mathrm{GE}
    = (2\pi)^3 \delta (\sum_a \vect k_a)
    \frac{ 4 H_*^6}{ \prod_a(2 k_a^3) }
    \cdot \sum_s \sum_{\substack{a<b \cr c<d}} \frac{1}{k^3_{ab}}
    \epsilon_{ij}^s(\vect k_{ab}) \epsilon_{lm}^s(\vect k_{cd})
    k_a^i k_b^j k_c^l k_d^m
    \cdot \mathcal{I}_{abcd} \;,
    \\
\end{eqnarray}
where the sum is only over different indices $a,b,c,d$,
and we have collected an overall symmetry factor of $4$ which takes into
account the exchanges $a \leftrightarrow b$ and $c
\leftrightarrow d$. Rewriting this summation somewhat more explicitly,
one arrives at the formula
\begin{eqnarray}
    \fl \nonumber
    \langle \dpp _{\bk_1}\dpp _{\bk_2}\dpp _{\bk_3}\dpp_{\bk_4}
    \rangle_*^\mathrm{GE}
    = (2\pi)^3 \delta (\sum_a \vect{k}_a) \frac{ 4 H_*^6}{ \prod_a(2
    k_a^3)} \nonumber
    \\
    \mbox{} \times
    \sum_s \Bigg[
        \frac{1}{k^3_{12}} \epsilon_{ij}^s(\vect k_{12})
        \epsilon_{lm}^s (\vect k_{34}) k_1^i k_2^j k_3^l k_4^m
        \cdot ({\cal I}_{1234}+{\cal I}_{3412}) \nonumber \\
        \qquad \mbox{} +
        \frac{1}{k^3_{13}} \epsilon_{ij}^s(\vect k_{13})
        \epsilon_{lm}^s (\vect k_{24}) k_1^i k_3^j k_2^l k_4^m
        \cdot
        ({\cal I}_{1324}+{\cal I}_{2413}) \nonumber \\
        \qquad \mbox{} +
        \frac{1}{k^3_{14}} \epsilon_{ij}^s(\vect k_{14})
        \epsilon_{lm}^s (\vect k_{23}) k_1^i k_4^j k_2^l k_3^m
        \cdot
        ({\cal I}_{1423}+{\cal I}_{2314}) \Big] \; .
    \label{3_lines}
\end{eqnarray}

The calculation of $\mathcal{I}_{1234}$ is extremely lengthy but
straightforward, and is given explicitly in~\ref{appendix:integral}.
There are divergent terms in the limit $\eta_\ast
\rightarrow 0$ together with a logarithmic dependence on the momenta,
but these peculiarities cancel between $\mathcal{I}_{1234}$
and $\mathcal{I}_{3412}$ to give an answer which is finite at late
times and depends rationally on the $k_a$,
\begin{eqnarray}
 && \fl  {\cal I}_{1234} + {\cal I}_{3412} =
\frac{k_1+k_2}{a_{34}^2}\Bigg[\frac12
(a_{34}+k_{12})(a_{34}^2-2b_{34})
+k_{12}^2 (k_3+k_4) \Bigg] + (1,2 \leftrightarrow 3,4)\nonumber \\
 &&  \!\!\!\!\!\!\!\!\!\!\!\!\!\!\!
 +\frac{k_1k_2}{k_t} \Bigg[\frac{b_{34}}{a_{34}} -
k_{12} + \frac{k_{12}}{a_{12}} \left(k_3k_4 -k_{12}
\frac{b_{34}}{a_{34}}\right) \left(\frac1{k_t} +\frac1{a_{12}}
\right)\Bigg]
+(1,2 \leftrightarrow 3,4) \nonumber \\
  &&  \!\!\!\!\!\!\!\!\!\!\!\!\!\!\! -
 \frac{ k_{12}}{a_{12} a_{34} k_t} \Bigg[ b_{12} b_{34} + 2
k_{12}^2 \big( \prod_a k_a \big) \left(\frac{1} {k_t^2}
+\frac{1}{a_{12} a_{34}} +\frac{k_{12}}{k_t a_{12} a_{34}} \right)
\Bigg]~, \label{eq:I-integral}
\end{eqnarray}
where we have used $k_{12}=k_{34}$ and we have defined
\begin{equation}
    \label{a_b_defined}
    a_{ab}\equiv k_a+k_b+k_{ab}\;, \qquad b_{ab} \equiv (k_a+k_b) k_{ab}
    +k_a k_b\;.
\end{equation}

To simplify Eq.~\eref{3_lines} further, let us work out the
polarization factors. To start, we will
express $\sum_s \epsilon_{ij}^s(\vect k_{12}) \epsilon_{lm}^s
(\vect k_{34}) k_1^i k_2^j k_3^l k_4^m$ in terms of the relative
angles between the $\vect k_a$ and $\vect k_{12}$. The
polarization tensors $\epsilon^s_{ij}$ can be rewritten as
\begin{equation}
    \ep_{ij}^+=
    \vect e_i \vect e_j - \vect {\bar e}_i\vect {\bar
    e}_j\;,
    \qquad \ep_{ij}^\times= \vect e_i \vect {\bar e}_j + \vect
    {\bar e}_i\vect e_j\;,
\end{equation}
where $\vect e$ and $\vect {\bar e}$ are orthogonal unit vectors
perpendicular to $\vect k_{12}$. It is convenient to
write the $\vect k_a$ in a spherical coordinate system having $\{
\vect e, \vect {\bar e}, \vect {\hat k}_{12} \}$ as basis (we have
denoted $ \vect{ \hat{k}} \equiv \vect k / k$). In this coordinate
system one obtains
\begin{equation}
\vect k_a = k_a(\sin \theta_{a} \cos \phi_{a},\sin \theta_{a}
    \sin    \phi_{a}, \cos \theta_a)\;,
\end{equation}
where $\cos \theta_a \equiv \vect {\hat k}_a \cdot \vect {\hat
k}_{12} $ and $\cos \phi_a \equiv \vect {\hat k}_a \cdot \vect e
$. Thus
\begin{eqnarray}
    \ep_{ij}^+ k_1^i k_2^j &=&  k_1 k_2 \sin \theta_1 \sin \theta_2
    \cos (\phi_{1} + \phi_{2}) \;, \\ \ep_{ij}^\times k_1^i k_2^j &=&  k_1 k_2
    \sin \theta_1 \sin \theta_2 \sin (\phi_{1}+\phi_2) \;,
\end{eqnarray}
with an analogous relation holding for $\ep_{ij}^+ k_3^i k_4^j$
and $\ep_{ij}^\times k_3^i k_4^j$. Since the projections of $\vect
k_1$ and $\vect k_2$ ($\vect k_3$ and $\vect k_4$) on the plane
orthogonal to $\vect k_{12}$ have the same amplitude but opposite
directions, then $k_2 \sin \theta_2 = k_1 \sin \theta_1$  and
$\phi_2=\phi_1+\pi$ ($k_4 \sin \theta_4 = k_3 \sin \theta_3$ and
$\phi_4=\phi_3+\pi$). Using these relations, one finally finds
\begin{equation}
    \sum_s \epsilon_{ij}^s(\vect k_{12}) \epsilon_{lm}^s (\vect
    k_{34}) k_1^i k_2^j k_3^l k_4^m = k_1^2 k_3^2 \sin^2 \theta_1
    \sin^2 \theta_3 \cos 2 \chi_{12,34} \;, \label{angles}
\end{equation}
where $\chi_{12,34} \equiv \phi_{1} - \phi_{3}$ is the angle
between the projections of $\vect k_1$ and $\vect k_3$ on the
plane orthogonal to $\vect k_{12}$. This is also the angle between
the two planes formed by $\{\vect k_1, \vect k_2 \}$ and $\{\vect
k_3, \vect k_4\}$.
This expression can easily be generalized to the polarization
factors in the last two lines of Eq.~\eref{3_lines}.

We now wish to convert our prediction for the scalar four-point
correlation function due to graviton exchange into a prediction
for the correlation function of the curvature perturbation,
$\zeta$. In order to do so, it is sufficient to use the linear
relation between $\zeta$ and the scalar field perturbation $\dpp$
at Hubble crossing, {\ie}, $\zeta = \dpp_*/\sqrt{2\epsilon}$. As
shown in Refs.~\cite{Seery:2006vu,Seery:2006js,Byrnes:2006vq},
higher order terms in the
relation between $\zeta$ and $\dpp_*$ generate contributions to
the trispectrum that are suppressed by slow-roll parameters with
respect to the contribution from graviton exchange.  Thus,
using Eq.~\eref{angles} and analogous expressions
for the other polarization factors, we can rewrite Eq.~\eref{3_lines} to give
\begin{eqnarray}
    \fl \nonumber
    \langle \zeta _{\bk_1}\zeta _{\bk_2}\zeta _{\bk_3}\zeta_{\bk_4}
    \rangle^\mathrm{GE}
    = (2\pi)^3 \delta (\sum_a \vect{k}_a) \frac{  H_*^6}{ \epsilon^2 \prod_a(2
    k_a^3)} \nonumber
    \\
    \mbox{} \times \Bigg[
        \frac{k_1^2 k_3^2}{k^3_{12}}  [1-(\hat {\vect k}_{1} \cdot \hat {\vect
k}_{12})^2] [1-(\hat {\vect k}_{3} \cdot \hat {\vect k}_{12})^2]
\cos 2\chi_{12,34}
        \cdot ({\cal I}_{1234}+{\cal I}_{3412}) \nonumber \\
         \ \, \mbox{} +
        \frac{k_1^2 k_2^2}{k^3_{13}} [1-(\hat {\vect k}_{1} \cdot \hat {\vect
k}_{13})^2] [1-(\hat {\vect k}_{2} \cdot \hat {\vect k}_{13})^2]
\cos 2\chi_{13,24} \cdot
        ({\cal I}_{1324}+{\cal I}_{2413}) \nonumber \\
         \ \, \mbox{} +
        \frac{k_1^2 k_2^2}{k^3_{14}} [1-(\hat {\vect k}_{1} \cdot \hat {\vect
k}_{14})^2] [1-(\hat {\vect k}_{2} \cdot \hat {\vect k}_{14})^2]
\cos 2\chi_{14,23} \cdot
        ({\cal I}_{1423}+{\cal I}_{2314}) \Big] \; . \nonumber \\
    \label{3_lines_final}
\end{eqnarray}

%%%%%%%%%%%%%%%%%%%%%%%%%%%%%%%%%%%%%%%%%%%%%%%%%%%%%%%%%%%%%%%%%%%%%%%%%%%%
%%%%%%%%%%%%%%%%%%%%%%%%%%%%%%%%%%%%%%%%%%%%%%%%%%%%%%%%%%%%%%%%%%%%%%%%%%%%

\section{The counter-collinear limit}
\label{sec:consistency}

%%%%%%%%%%%%%%%%%%%%%%%%%%%%%%%%%%%%%%%%%%%%%%%%%%%%%%%%%%%%%%%%%%%%%%%%%%%%
%%%%%%%%%%%%%%%%%%%%%%%%%%%%%%%%%%%%%%%%%%%%%%%%%%%%%%%%%%%%%%%%%%%%%%%%%%%%

The calculation of the four-point correlation induced by the
graviton exchange process is rather complicated, and the final
expression given in Eq.~\eref{3_lines_final}---combined with
Eq.~\eref{eq:I-integral}---is not simple to interpret. One would
therefore like an independent check that the result is correct.
It is possible to achieve this in the counter-collinear
limit, where
$\vect k_{ab} = \vect{k}_a + \vect{k}_b \rightarrow 0$
and we can attach a simple physical meaning to the exchange
process: in this limit, the four-point function simply expresses
the correlation between a pair of two-point functions, for
instance $\langle \zeta_{\vect{k}_1} \zeta_{\vect{k}_2} \rangle$
and $\langle \zeta_{\vect{k}_3} \zeta_{\vect{k}_4} \rangle$
which is induced by a low frequency gravitational wave of
momentum $k_{12} \ll k_1 \approx k_2, k_3
\approx  k_4$. This gravitational wave crossed the Hubble radius
much earlier than any of the $k_a$ modes and can be considered as
a fixed classical background into which they exit \cite{Maldacena:2002vr}.
As discussed in
\S\ref{sec:introduction}, in this limit the momentum quadrilateral
constitutes a planar parallelogram (or a folded-kite).

Exactly the same reasoning applies where the exchanged particle is a scalar
mode, and allows us to estimate the size of the contribution from such a
process in the counter-collinear limit. Following Ref.~\cite{Huang:2006eh}
(see also Ref.~\cite{Cheung:2007sv}), we write the background mode as
$\zeta_B(\vect k_0)$. The effect of this mode is to re-scale the
spatial coordinates so that the two-point function $\langle
\zeta_{\bk_1} \zeta_{\bk_2} \rangle$ receives a correction
proportional to its tilt,
\begin{equation}
    \fl
    \langle \zeta_{\bk_1} \zeta_{\bk_2} \rangle_{\zeta_B}
    =
    \langle \zeta_{\bk_1} \zeta_{\bk_2} \rangle
    - (n_s -1) \int \frac{d^3 k_0}{(2\pi)^3}
        \zeta_B(\bk_0) (2\pi)^3 \delta (\bk_1+\bk_2-\bk_0)
        P_\zeta(k_2)\;.
\end{equation}
A similar formula gives the transformation of
$\langle \zeta_{\vect k_3} \zeta_{\vect k_4} \rangle$. To obtain
the contribution to the four-point function due to the scalar
background mode (which we label with the superscript ``SE'') in
the counter-collinear limit, it follows that we need only to
average over this pair of two-point functions, which yields
\begin{equation}
   \fl \langle
        \langle \zeta_{\bk_1} \zeta_{\bk_2} \rangle_{\zeta_B}
        \langle \zeta_{\bk_3} \zeta_{\bk_4} \rangle_{\zeta_B}
    \rangle^\mathrm{SE} =
    (2 \pi)^3 \delta (\sum \vect k_a) \cdot (n_s-1)^2  \cdot  P_\zeta(k_{12})
    P_\zeta(k_1) P_\zeta(k_3)\;.
	\label{eq:scalar-squeezed}
\end{equation}
Thus, in the counter-collinear limit,
the correlations induced by scalar exchange are small, being
suppressed by two powers of the scalar tilt. This is consistent
with the estimate made in the Introduction
(\S\ref{sec:introduction}), which found the contribution from
scalar exchange to be suppressed by $\Or(\epsilon^2, \eta^2,
\epsilon \eta)$.
Note that Eq.~\eref{eq:scalar-squeezed}
can be thought of as arising from the product
of two back-to-back bispectra in the squeezed limit, and exactly
satisfies the classical relationship between $\tnl$ and $\fnl$
described in Eq.~\eref{eq:local-trispectrum}. Nevertheless, the full 
trispectrum does not. Indeed, as we show below, although the contributions
from the contact interactions are subdominant in this limit
\cite{Seery:2006vu},
those from gravitational wave exchange remain important
and dominate the trispectrum.

Now consider the effect of a long wavelength gravitational wave on
$\langle \zeta_{\bk_1} \zeta_{\bk_2} \rangle$. This will turn out
to be much larger than the effect due to the scalar mode. In this
case the effect of the background gravitational wave,
$\gamma_B^{ij}(\vect k_0)$,
is to deform the geometry in such a way that a small spatial
distance $d x^2$ is rescaled to $dx^2 \to (e^\gamma)_{ij}dx^idx^j
\simeq dx^2+\gamma_{ij}dx^idx^j$. In Fourier space we have
\begin{equation}
    \fl
    \langle \zeta_{\bk_1} \zeta_{\bk_2} \rangle_{\gamma_B}
    = \langle \zeta_{\bk_1} \zeta_{\bk_2} \rangle
    + \int \frac{d^3 k_0}{(2\pi)^3}
    \gamma_B^{ij}(\bk_0) \left( \frac{\partial}{\partial \gamma^{ij}_B}
    \langle \zeta_{\tilde \bk_1} \zeta_{\tilde \bk_2} \rangle_{\gamma_B}
    \right)_{\gamma_B=0}\;,
\end{equation}
where $\tilde \bk_a = \bk_a-\bk_0/2$. The effect of the background
tensor mode is to deform the momenta such that $k^2 \to
k^2-\gamma_{ij}k^i k^j$ \cite{Maldacena:2002vr}, so that the
equation above becomes
\begin{eqnarray}
    \fl\nonumber
    \langle \zeta_{\bk_1} \zeta_{\bk_2} \rangle_{\gamma_B}
    = \langle \zeta_{\bk_1} \zeta_{\bk_2} \rangle -
    \int \frac{d^3 k_0}{(2\pi)^3} \gamma_B^{ij}(\bk_0) k^1_i
	k^1_j
    \left( \frac{\partial}{\partial (k^1)^2}
        \langle \zeta_{\tilde \bk_1} \zeta_{\tilde \bk_2} \rangle_{\gamma_B}
    \right)_{\gamma_B=0}\\
    = \langle \zeta_{\bk_1} \zeta_{\bk_2} \rangle -
    \int \frac{d^3 k_0}{(2\pi)^3} (2 \pi)^3 \delta(\bk_1 +\bk_2 -\bk_0)
    \gamma_B^{ij}(\bk_0) \frac{k^1_i k^1_j}{k_1^2} \frac{3}{2} P_\zeta(k_1)\;.
    \label{eq:graviton-exchange}
\end{eqnarray}
In this case, the effect of deforming the geometry is not
suppressed by the scalar tilt, and therefore dominates compared
with the effect of a background scalar mode. Nevertheless,
Eq.~\eref{eq:graviton-exchange} is proportional to the amplitude
of gravitational waves, which is of order $\sim r^{-1/2}$
when compared with the scalar amplitude. For this reason, one
finds that the trispectrum in this limit is proportional to the
ratio of the tensor to scalar power spectra.

Indeed, taking the average of $\langle \zeta_{\vect k_1}
\zeta_{\vect k_2} \rangle_{\gamma_B}$ and $\langle \zeta_{\vect
k_3} \zeta_{\vect k_4} \rangle_{\gamma_B}$ we find the relation
\begin{eqnarray}
    \fl \nonumber
    \langle \langle \zeta_{\vect k_1} \zeta_{\vect k_2}
    \rangle_{\gamma_B} \langle \zeta_{\vect k_3} \zeta_{\vect k_4}
    \rangle_{\gamma_B} \rangle^\mathrm{GE} =
    (2\pi)^3
    \delta (\sum_a \bk_a)
    \\
    \mbox{} \times
    \sum_s \epsilon^s_{ij} (\vect k_{12}) \frac{k_1^i k^j_1}{k_1^2}
    \epsilon^s_{lm} (\vect k_{34}) \frac{k_3^l k^m_3}{k_3^2}
    \cdot 4 \ep \cdot \frac{9}{4}
    \cdot P_\zeta(k_{12}) P_\zeta(k_1) P_\zeta(k_3)\;.
    \label{cc_limit}
\end{eqnarray}
We can use Eq.~\eref{angles} to rewrite the polarization factor in
this equation. In the limit $k_{12} \to 0$, $\theta_1, \theta_3
\to \pi$ and Eq.~\eref{angles} reads
\begin{equation}
    \sum_s \epsilon^s_{ij} (\vect k_{12}) {k_1^i k^j_1}
    \epsilon^s_{lm} (\vect k_{34}) {k_3^l k^m_3}
    = k_1^2 k_3^2
    \cos 2 \chi_{12,34}\;.
\end{equation}
Thus, Eq.~\eref{cc_limit} yields a semiclassical relation in the
counter-collinear limit ($k_{12} \ll k_1 \approx k_2, k_3 \approx
k_4$),
\begin{equation}
    \fl
    \langle
        \zeta_{\vect k_1} \zeta_{\vect k_2} \zeta_{\vect k_3} \zeta_{\vect k_4}
    \rangle^\mathrm{GE} = (2\pi)^3 \delta (\sum_a \vect k_a)
    \, \cdot \left( \frac{9}{16} r
    	\cos 2 \chi_{12,34} \right) \cdot
    P_\zeta(k_1) P_\zeta(k_3) P_\zeta(k_{12})
        \;,
    \label{limit12}
\end{equation}
where we have used $r =16 \epsilon$ and
in this degenerate limit $\chi_{12,34}$ becomes the angle between
$\bk_1$ and $\bk_3$.

Let us check that Eq.~\eref{3_lines_final} satisfies this
relation and consistently reproduces Eq.~\eref{limit12} in the
counter-collinear limit. Indeed, for $k_{12} \ll k_1 \approx k_2,
k_3 \approx k_4$
the second line of Eq.~\eref{3_lines_final} dominates over the
last two and we must only compute the sum of the integrals ${\cal
I}_{1234}+{\cal I}_{3412}$ by using Eq.~\eref{eq:I-integral} in
this limit. For $k_{12} = 0$, $k_1=k_2$ and $k_3=k_4$. Using
Eq.~\eref{a_b_defined}, $a_{12}=2 k_1$, $a_{34}=2k_3$,
$b_{12}=k_1^2$ and $b_{34}=k_3^2$. Thus, the right hand side of
the first line of Eq.~\eref{eq:I-integral} yields $2 k_1 k_3 $,
the second line yields $k_1 k_3/4$ while the third line vanishes.
Then, one finds ${\cal I}_{1234}+{\cal I}_{3412} = \frac{9}{4} k_1
k_3 $ and Eq.~\eref{3_lines_final} becomes
\begin{equation}
    \langle
        \zeta_{\vect k_1} \zeta_{\vect k_2} \zeta_{\vect k_3} \zeta_{\vect k_4}
    \rangle^\mathrm{GE} = (2\pi)^3 \delta (\sum_a \vect k_a)
    \frac{ H_*^6}{\epsilon^2 2^4 k_1^3 k_3^3 k_{12}^3}
    \cdot \frac{9}{4} \cos 2\chi_{12,34}   \;.
\end{equation}
Comparing this equation with~\eref{limit12} after using
the relations $P_\zeta(k) = 
H_*^2/(4\epsilon k^3)$ and $r=16 \epsilon$,
one sees that the semiclassical relation is satisfied.

Note that, strictly speaking, Eq.~\eref{limit12} is \emph{not} a
consistency relation like Maldacena's consistency relation in
the squeezed limit. In particular, the full trispectrum need not
satisfy this relation. This relation will be satisfied by the full
trispectrum only when contact interactions are negligible with
respect to the graviton exchange in the counter-collinear limit.
Accidentally, as we will show below, this is the case for
slow-roll inflation, and the full trispectrum satisfies this
relation. However, this is not the case, for instance, in models
with non-standard kinetic terms.

%%%%%%%%%%%%%%%%%%%%%%%%%%%%%%%%%%%%%%%%%%%%%%%%%%%%%%%%%%%%%%%%%%%%%%%%%%%%
%%%%%%%%%%%%%%%%%%%%%%%%%%%%%%%%%%%%%%%%%%%%%%%%%%%%%%%%%%%%%%%%%%%%%%%%%%%%

\section{The total trispectrum}
\label{sec:tot}

%%%%%%%%%%%%%%%%%%%%%%%%%%%%%%%%%%%%%%%%%%%%%%%%%%%%%%%%%%%%%%%%%%%%%%%%%%%%
%%%%%%%%%%%%%%%%%%%%%%%%%%%%%%%%%%%%%%%%%%%%%%%%%%%%%%%%%%%%%%%%%%%%%%%%%%%%

The connected four-point correlation function of field
perturbations at horizon crossing was computed in
Ref.~\cite{Seery:2006vu}, assuming that inflation was driven by a
collection of light scalar fields with canonically normalized
kinetic terms. However, this calculation was restricted only to
contact terms, neglecting correlations induced by particle
exchange. In this section we put together the trispectrum of
Ref.~\cite{Seery:2006vu} for a single field with the exchange
contribution computed in \S\ref{sec:trispectrum}.
This gives the full trispectrum to leading order in slow-roll.
Then we discuss this result.

When specialized to the case of a single field, the contact
interaction (labelled with a superscript ``CI'')
gives a contribution to the scalar trispectrum equal
to \cite{Seery:2006vu}
\begin{equation}
 \fl\langle
    \zeta_{\bk_1} \zeta_{\bk_2}\zeta_{\bk_3}\zeta_{\bk_4}
    \rangle^\mathrm{CI}
    = (2 \pi)^3 \delta(\sum_a \vect k_a)
    \frac{ H_*^6}{4 \epsilon^2 \prod_a (2k^3_a)} \sum_{\mathrm{perms}}
    {\cal M}_4(\vect k_1,\vect k_2,\vect k_3,\vect k_4)\;,
    \label{tri_sls}
\end{equation}
where we have defined a form-factor $\mathcal{M}_4$ which encapsulates the
detailed momentum dependence. This is given by the expression
\begin{eqnarray}
    \fl\nonumber
    {\cal M}_4(\vect k_1,\vect k_2,\vect k_3,\vect k_4)
    = - 2
    \frac{k_1^2 k_3^2 }{k_{12}^2 k_{34}^2}
    \frac{W_{24}}{k_t} \left(  \frac{\bZ_{12} \cdot
    \bZ_{34}}{k_{34}^2} + 2 \vect k_2 \cdot \bZ_{34} +
    \frac{3}{4}\sigma_{12} \sigma_{34}  \right)
    \\ \mbox{} -
    \frac{1}{2} \frac{k_3^2}{k_{34}^2} \sigma_{34} \left(
    \frac{\bk_1 \cdot \bk_2}{k_t} W_{124} + 2 \frac{k_1^2 k_2^2}{k_t^3}
    + 6 \frac{k_1^2 k_2^2 k_4}{k_t^4} \right)\;, \label{form_factor}
\end{eqnarray}
where the quantities occurring here are defined by
\begin{eqnarray}
    \sigma_{ab} & = \bk_a \cdot \bk_b + k_b^2\;, \\
    \bZ_{ab} & = \sigma_{ab} \bk_a - \sigma_{ba} \bk_b\;, \\
    W_{ab} & = 1+ \frac{k_a+k_b}{k_t}+ \frac{2 k_a k_b}{k_t^2}\;, \\
    W_{abc} & = 1+ \frac{k_a+k_b+k_c}{k_t} +
    \frac{2 (k_a k_b + k_b k_c + k_a k_c)}{k_t^2} +
    \frac{6 k_a k_b k_c}{k_t^3}\;.
\end{eqnarray}

In order to write Eq.~\eref{tri_sls} from
Ref.~\cite{Seery:2006vu}, we have used the linear relation $\zeta
= \dpp_*/\sqrt{2\epsilon}$ to express $\zeta$ in terms of the
field perturbations at Hubble crossing $\dpp_*$.
As discussed earlier, nonlinearities in
the relation between $\zeta$ and $\dpp_*$ lead to corrections to
the trispectrum that are suppressed by slow-roll with respect
to~\eref{tri_sls} \cite{Seery:2006vu,Seery:2006js,Byrnes:2006vq}.
Thus, at leading order in slow-roll,
the total trispectrum is given by the sum of the two
contributions, Eqs.~\eref{3_lines_final} and~\eref{tri_sls},
\begin{equation}
    \label{main}
    \langle \zeta_{\bk_1} \zeta_{\bk_2}\zeta_{\bk_3}\zeta_{\bk_4}
    \rangle
    =
    \langle
        \zeta_{\bk_1} \zeta_{\bk_2}\zeta_{\bk_3}\zeta_{\bk_4}
    \rangle_*^\mathrm{CI} +
    \langle
        \zeta_{\bk_1} \zeta_{\bk_2}\zeta_{\bk_3}\zeta_{\bk_4}
    \rangle_*^\mathrm{GE}
    \;.
\end{equation}

We have already discussed the squeezed limit, in which
one of the momenta becomes much smaller than the others. In this limit,
the trispectrum must be described by Maldacena's
consistency relation
\cite{Maldacena:2002vr,Creminelli:2004yq,Huang:2006eh,
Chen:2006nt,Cheung:2007sv}:
the scalar mode associated with the small momentum forms a fixed
classical background into which the remaining modes exit, in a
similar fashion to the calculations discussed in
\S\ref{sec:consistency}. In Ref.~\cite{Seery:2006vu} it was noted
that the scalar contact term is one order \emph{lower} in
slow-roll than the term yielded by the consistency relation in
this limit. These two apparently contradictory statements are
nevertheless in agreement, because the leading-order slow-roll
piece in the contact term vanishes in the squeezed limit. Although
we have not pursued the calculation of the trispectrum to a
sufficiently high order in the slow-roll to verify that
Maldacena's relation holds in detail for the graviton exchange process,
our final result for the trispectrum,
Eq.~\eref{main},
is also one order lower in slow-roll than predicted by the
consistency relation, but vanishes in the squeezed limit. It is
therefore consistent with Maldacena's result.

In the counter-collinear limit
$k_{12} \ll k_1 \approx k_2, k_3 \approx k_4$
the form-factor $\mathcal{M}_4$ defined in Eq.~\eref{form_factor}
remains finite~\cite{Seery:2006vu}, so that the total trispectrum
in Eq.~\eref{main} is dominated by the contribution from graviton
exchange. One can show that in this limit it is possible to
parameterize the total trispectrum~\eref{main} in terms of the
local parameter
$\tnl^\mathrm{local}$. Indeed, neglecting the contribution from
contact interactions, the total trispectrum
is given by Eq.~\eref{limit12}. Thus, comparing this equation with
Eq.~\eref{local_2} gives
\begin{equation}
    \tnl^\mathrm{local}
    =\frac{9}{64} r
    \cos 2 \chi_{12,34}\;.
    \label{tnl_folded_kite}
\end{equation}
Thus, in the counter-collinear limit,
the trispectrum of single-field slow-roll inflation is of the
local type in real space, with a magnitude expressed by
Eq.~\eref{tnl_folded_kite}. This is consistent with our
expectations: indeed, as was discussed in the Introduction
({\S}\ref{sec:introduction}), we
expect classical processes operating outside the horizon to
synthesize local contributions to the curvature perturbation.
The counter-collinear limit of graviton exchange is tantamount to
such a process: it corresponds
to a gravitational wave which nucleates inside the horizon in the
far past of inflation, and subsequently passes beyond the Hubble radius
where it becomes approximately classical. In the diagrammatic
interpretation, this extremely soft graviton then splits into four
scalar quanta which propagate to their own hypersurface of horizon
exit. These four quanta are correlated because they all originate
from the same parent graviton, or gravitational wave.

%%%%%%%%%%%%%%%%%%%%%%%%%%%%%%%%%%%%%%%%%%%%%%%%%%%%%%%%%%%%%%%%%%%%%%%%%%%%
%%%%%%%%%%%%%%%%%%%%%%%%%%%%%%%%%%%%%%%%%%%%%%%%%%%%%%%%%%%%%%%%%%%%%%%%%%%%

\section{Discussion and conclusion}
\label{sec:conclude}

%%%%%%%%%%%%%%%%%%%%%%%%%%%%%%%%%%%%%%%%%%%%%%%%%%%%%%%%%%%%%%%%%%%%%%%%%%%%
%%%%%%%%%%%%%%%%%%%%%%%%%%%%%%%%%%%%%%%%%%%%%%%%%%%%%%%%%%%%%%%%%%%%%%%%%%%%

We have calculated a new contribution to the inflationary
trispectrum, arising from a process which can be visualized as
graviton exchange. This contribution is of the same order of
magnitude as the one previously calculated by considering only
scalar and vector perturbations and gives rise to an effect of
magnitude $\tnl \sim \epsilon$. Although one might have expected
$\tnl \sim \Or(\epsilon^2, \eta^2, \epsilon \eta)$ on the basis of
Maldacena's consistency relation, we can conclude that the
estimate $\tnl \sim \epsilon$ made in Ref.~\cite{Seery:2006vu} is
essentially correct, although modified in detail by numerical
factors coming from the exchange process which was neglected in
that work.
Note that although we have framed our calculations in the context of
single-field inflation, our results for the curvature perturbation
also hold for models with multiple
fields and a flat field space metric.

We would like to emphasize that the relation $\tnl \sim \fnl^2$
[see Eq.~\eref{local_relation}], which one might anticipate based
on comparison of the local formulae~\eref{eq:local-bispectrum}
and~\eref{eq:local-trispectrum}, need not apply. Indeed, it does
not apply in single-field slow-roll inflation models. As explained
in the Introduction, one can only trust such classical
relationships if the non-Gaussianity is generated by
gravitational evolution subsequent to horizon crossing. In
single-field inflation, non-Gaussianities are a quantum
interference effect and are produced at Hubble crossing, after
which the curvature perturbation $\zeta$ remains constant.

In Ref.~\cite{Seery:2006vu}, it was supposed that the effect from
quantum processes around horizon crossing would be maximized on a
configuration close to the equilateral case, where all $k_a$ take
the same magnitude $k$. In the exactly equilateral limit $|\tnl|$
can be computed analytically, giving a maximum effect which
corresponds to
\begin{equation}
    \tnl^{\mathrm{CI}} \leq \frac{23}{576 \sqrt{3}} r \approx 0.0231 r
    \approx \frac{r}{44}
    \hfill \mbox{(equilateral and $\gnl =0$)},
	\hspace{5mm}
\end{equation}
where
to compute $\tnl$ from the trispectrum
we have set $\gnl=0$. The tensor to scalar ratio $r$ is
currently subject to the observational constraint $r < 0.22$ at 95\% CL
from a combination of the WMAP five-year, baryon oscillation and
supernova data \cite{Komatsu:2008hk}. (This bound was approximated
as $\tnl \lesssim r/50$
in Ref.~\cite{Seery:2006vu}.) By searching
through the available parameter space, we have subsequently
located a marginally off-equilateral configuration which gives a
non-linearity of greater magnitude, but opposite sign,
\begin{equation}
    \tnl^{\mathrm{CI}} = 0.0298 r \approx \frac{r}{34}
    \hfill \mbox{(side ratios 1\,:\,1\,:\,1\,:\,2.96 and $\gnl =0$)} .
	\hspace{5mm}
    \label{eq:largest-ci}
\end{equation}

How large a correlation can be induced by the effect of graviton
exchange?
It is difficult to pursue analytic estimates for the graviton
exchange contribution. Proceeding numerically, however, it is
possible to find a marginally squeezed configuration which gives
an effect corresponding to
\begin{equation}
    \tnl^{\mathrm{GE}} = -1.217 r
    \hfill \mbox{(side ratios 1\,:\,10\,:\,10\,:\,13.44 and $\gnl =0$)} .
	\hspace{5mm}
    \label{eq:largest-ge}
\end{equation}
From Eqs.~\eref{eq:largest-ci} and~\eref{eq:largest-ge}, it is clear that
the total trispectrum will be dominated by the contribution from
graviton exchange. Taking both effects together, we have located
a configuration which gives a combined $\tnl$ equivalent to
\begin{equation}
    \tnl = -1.218 r
    \hfill \mbox{(side ratios 1\,:\,5.6\,:\,10\,:\,10 and $\gnl =0$)} .
	\hspace{5mm}
\end{equation}
We would like to emphasize, however, that our numerical search has
not been exhaustive, and it is possible that other configurations
exist which give even larger $|\tnl|$.
These estimates hold for multiple-field inflation provided that the
hierarchy among the sides of the momentum quadrilateral is not
too large \cite{Seery:2006vu}.
For instance, the ratio of sides given in Eq.~\eref{eq:largest-ge}
corresponds to
an interval of $\sim 2.6$ e-folds between
horizon exit of the modes corresponding to largest and smallest wavelengths.
For such a small interval, we can expect that the approximation of
almost simultaneous horizon exit is reasonably satisfied.
On the other hand, in the
limit where a large hierarchy exists among the ratio of sides in
the quadrilateral, our analytical expression becomes untrustworthy.

We can therefore conclude that the trispectrum at horizon crossing is generally
dominated by the contribution from graviton exchange, and that
at maximum the total trispectrum is roughly of order $|\tnl| \sim r$.

\section*{Acknowledgments}

It is a pleasure to acknowledge helpful discussions with
Frederico Arroja, Paolo Creminelli, Kazuya Koyama
and Marcello Musso.
We would like to thank Peter Adshead, Richard Easther and Eugene Lim
for very interesting correspondence, and for spotting several
typos in earlier versions of this paper.

FV would like to
thank the Department of Physics and Astronomy of the University of
Aarhus, where this work was initiated, for their kind hospitality.
Furthermore, we would like to thank the Department of Applied
Mathematics and Theoretical Physics at Cambridge for their
hospitality while this work was being carried out---during the
workshop \emph{Non-Gaussianity from fundamental physics} held in
September 2008---and the EU Marie Curie Research \& Training
network ``UniverseNet'' (MRTN-CT-2006-035863) for support.
MSS would also like to thank the Institute of Theoretical Astrophysics
in Oslo for their hospitality during Fall 2008. DS is supported by STFC.

%%%%%%%%%%%%%%%%%%%%%%%%%%%%%%%%%%%%%%%%%%%%%%%%%%%%%%%%%%%%%%%%%%%%%%%%%%%%
%%%%%%%%%%%%%%%%%%%%%%%%%%%%%%%%%%%%%%%%%%%%%%%%%%%%%%%%%%%%%%%%%%%%%%%%%%%%

\appendix

%%%%%%%%%%%%%%%%%%%%%%%%%%%%%%%%%%%%%%%%%%%%%%%%%%%%%%%%%%%%%%%%%%%%%%%%%%%%
%%%%%%%%%%%%%%%%%%%%%%%%%%%%%%%%%%%%%%%%%%%%%%%%%%%%%%%%%%%%%%%%%%%%%%%%%%%%

\section{Computing the time integral}
\label{appendix:integral}

%%%%%%%%%%%%%%%%%%%%%%%%%%%%%%%%%%%%%%%%%%%%%%%%%%%%%%%%%%%%%%%%%%%%%%%%%%%%
%%%%%%%%%%%%%%%%%%%%%%%%%%%%%%%%%%%%%%%%%%%%%%%%%%%%%%%%%%%%%%%%%%%%%%%%%%%%

In this Appendix we compute the sum of the time integrals ${\cal
I}_{1234} + {\cal I}_{3412}$. Let us start by computing
${\cal I}_{1234}$, which reads
\begin{eqnarray}
    \fl\nonumber
%    \label{int_1234}
    {\cal I}_{1234} \equiv
    \int^{\eta_*}_{-\infty} \frac{d\eta}{\eta^2} \int^\eta_{-\infty}
    \frac{d\eta'}{\eta'^2}
    \Im
        \Big  [ (1+ik_1 \eta_*) (1+ik_2 \eta_*)
                (1-ik_1 \eta) (1-ik_2 \eta) e^{i(\eta-\eta_\ast)(k_1+k_2)}
        \Big  ]
    \\ \nonumber
    \mbox{} \times
    \Im
        \Big  [ (1+ik_3 \eta_*) (1+ik_4 \eta_*)
                (1-ik_3 \eta')(1-ik_4 \eta')
                \\ \qquad\quad \mbox{} \times
                (1+ik_{12} \eta)
                (1-ik_{12} \eta') e^{i(\eta'-\eta_\ast)(k_3+k_4)}
                e^{-i(\eta-\eta')k_{12}}
        \Big  ] \;,
\end{eqnarray}
It is convenient to divide this into the sum of 4 integrals.
Using the rule $  \Im(AB) =-\im (A B^\ast - A^\ast B)/2  $ we
obtain
\begin{equation}
{\cal I}_{1234} = -\frac{1}{4} \left(I_A -I_B-I_C+I_D\right)\;,
\label{eq:sum}
\end{equation}
where
\begin{eqnarray}
I_A \equiv \int^{\eta_*}_{-\infty} \frac{d\eta}{\eta^2} XY
\int^\eta_{-\infty}
\frac{d\eta'}{\eta'^2} Z\;, \\
I_B \equiv \int^{\eta_*}_{-\infty} \frac{d\eta}{\eta^2} X^*Y
\int^\eta_{-\infty}
\frac{d\eta'}{\eta'^2} Z\;, \\
I_C \equiv \int^{\eta_*}_{-\infty} \frac{d\eta}{\eta^2} XY^*
\int^\eta_{-\infty}
\frac{d\eta'}{\eta'^2} Z^*\;, \\
I_D \equiv \int^{\eta_*}_{-\infty} \frac{d\eta}{\eta^2} X^*Y^*
\int^\eta_{-\infty} \frac{d\eta'}{\eta'^2} Z^*\;,
\end{eqnarray}
and
\begin{eqnarray}
\fl X(\eta,\eta_*)\equiv(1+ik_1 \eta_*)(1+ik_2
\eta_*) (1-ik_1 \eta)(1-ik_2 \eta) e^{i(\eta-\eta_*)(k_1+k_2)}\;\\
\fl Y(\eta,\eta_*)\equiv(1+ik_{12} \eta) e^{-i\eta k_{12}}\;,\\
\fl Z(\eta',\eta_*)\equiv(1+ik_3 \eta_*)(1+ik_4 \eta_*)(1-ik_3
\eta')(1-ik_4 \eta') (1-ik_{12} \eta')
e^{i(\eta'-\eta_*)(k_3+k_4)+i\eta' k_{12}} \;.\nonumber \\
\end{eqnarray}
Each of these have to be solved separately.

Let us first consider $I_A$. Expanding the
$\eta_*$-dependent part near $\eta_* \to 0$, this can be written in the form
\begin{eqnarray}
    \nonumber
    I_A&=& \left(1+i \delta \eta_* -\frac{\delta^2}{2}\eta_*^2
    + \frac{1}{2} \sum_a k_a^2\eta_*^2 +\ldots \right) e^{-i\eta_*
    \delta}\\
    &\times&\int_{-\infty}^{\eta_\ast} \frac{\d \eta}{\eta^2}
    (1 - \im \alpha_1 \eta + \alpha_2 \eta^2 - \im \alpha_3 \eta^3 )
    \e{\im \eta \alpha_1}\nonumber \\
    &\times&
    \int_{-\infty}^{\eta} \frac{\d \eta'}{\eta'^2}
    (1 - \im \beta_1 \eta' + \beta_2 \eta'^2 - \im \beta_3 \eta'^3 )
    \e{\im \eta' \beta_1}\;,
    \label{eq:Isymbol}
\end{eqnarray}
where
\begin{eqnarray}
\alpha_1\equiv k_1+k_2-k_{12}\;,\\
\alpha_2\equiv (k_1+k_2)k_{12}-k_1 k_2\;,\\
\alpha_3\equiv k_1k_2k_{12}\;,\\
\beta_1\equiv k_3+k_4+k_{12}\;,\\
\beta_2\equiv -(k_3+k_4)k_{12}-k_3 k_4\;,\\
\beta_3\equiv -k_3k_4k_{12}\;,\\
\delta\equiv \alpha_1+\beta_1\;.
\end{eqnarray}
The inner integral in the last line of Eq.~\eref{eq:Isymbol} can
be easily solved and its log divergent part cancels out. Plugging
its solution into Eq.~\eref{eq:Isymbol} yields
\begin{eqnarray}
  \fl\nonumber
  I_A = \left(1+i \delta \eta_* -\frac{\delta^2}{2}\eta_*^2
    + \frac{1}{2} \sum_a k_a^2\eta_*^2 +\ldots \right) e^{-i\eta_*
    \delta} \\
   \mbox{} \times \int_{-\infty}^{\eta_\ast} \d \eta \;
    \left( - \frac{1}{\eta^3} + \frac{\im \gamma_1}{\eta^2}
        + \frac{\gamma_2}{\eta}
        + \im \gamma_3
        + \gamma_4 \eta
        + \im \gamma_5 \eta^2
    \right) \e{\im \eta \delta} ,
\end{eqnarray}
where the $\gamma$'s are defined by
\begin{eqnarray}
    \gamma_1 & \equiv \alpha_1 - \frac{\beta_1 \beta_2 + \beta_3}{\beta_1^2}\;, \\
    \gamma_2 & \equiv - \frac{\alpha_1}{\beta_1^2}(\beta_1 \beta_2 + \beta_3) -
        \alpha_2 - \frac{\beta_3}{\beta_1}\;, \\
    \gamma_3 & \equiv \alpha_3 - \frac{\alpha_2}{\beta_1^2}(\beta_1 \beta_2 + \beta_3)
        + \frac{\alpha_1 \beta_3}{\beta_1}\;, \\
    \gamma_4 & \equiv - \frac{\alpha_3}{\beta_1^2}(\beta_1 \beta_2 + \beta_3)
        - \frac{\alpha_2 \beta_3}{\beta_1}\;, \\
    \gamma_5 & \equiv \frac{\alpha_3 \beta_3}{\beta_1} .
\end{eqnarray}
Carrying out the remaining $\eta$ integration, the final result is
\begin{eqnarray}
    \fl\nonumber
    I_A =
    \delta \left( \gamma_1 - \frac{3}{4} \delta \right) +
    \frac{\gamma_3}{\delta} + \frac{\gamma_4}{\delta^2} -2
    \frac{\gamma_5}{\delta^3}
    \\  \mbox{} + \frac{1}{2\eta_\ast^2} +\frac{1}{4} \sum_ak_a^2 + \frac{\im}{\eta_\ast} (\delta -
\gamma_1) -
    \left( \gamma_2 - \delta\gamma_1 + \frac{\delta^2}{2}  \right)
    E_1(-\im \delta \eta_\ast) , \label{eq:final_int}
\end{eqnarray}
where $E_1$ is the exponential integral or incomplete $\Gamma$
function,
\begin{equation}
    E_1(z) = \int_{z}^{\infty} \frac{\e{-t}}{t} \; \d t .
\end{equation}
This calculation can be repeated for the remaining integrals
$I_{B,C,D}$. The final result~\eref{eq:final_int} can be
generalized to $I_B$ by letting $k_1 \rightarrow - k_1$ and
$k_2 \rightarrow - k_2$;
to $I_C$ by the replacements $k_3 \rightarrow - k_3$,
$k_4 \to - k_4$ and $k_{12} \rightarrow -k_{12}$;
and to $I_D$ by allowing all the $k$s to change sign.
Eq.~\eref{eq:sum} can then be extended to ${\cal I}_{3412}$ by
exchanging $12 \leftrightarrow 34$.

Note that in the last line of Eq.~\eref{eq:final_int} there are
terms that diverge for $\eta_* \to 0$. However, we will now show
that this line does not contribute to the final combination ${\cal
I}_{1234} + {\cal I}_{3412}$ and can be thus dropped from the
discussion. Indeed, the first and second terms in this line are the
same for all the four integrals $I_{A,B,C,D}$. Thus they do not
contribute to the sum in Eq.~\eref{eq:sum}. The third term of
this line is odd under exchange of $I_A \leftrightarrow I_D$ and
$I_B \leftrightarrow I_C$. Thus, this term also does not
contribute to the sum~\eref{eq:sum}. Furthermore, it is
straightforward to show that the last term of this line does not
contribute when considering the final sum ${\cal I}_{1234} + {\cal
I}_{3412}$.
One might be concerned that
Eq.~\eref{eq:final_int} contains apparently divergent
(and unphysical) terms in the limit
$\delta \rightarrow 0$, which occurs when the momentum quadrilateral
becomes counter-collinear. Although it is not possible to see cancellation
at intermediate stages in the calculation, these terms also cancel in
${\cal I}_{1234} + {\cal I}_{3412}$.%
	\footnote{\emph{Note added after publication}.
	We have chosen to nest the terms arising from $(+,-)$ 
	and $(-,+)$ diagrams, which enter with a minus sign inside 
	the square bracket of Eq.~\eref{int_1234}.
	The right hand side of Eq.~\eref{int_1234}
	could alternatively have been evaluated by 
	factorization, which leads to an expression of the form
	\begin{eqnarray}
		\fl
	     \langle \dpp_{{\bf k}_1} \dpp_{{\bf k}_2}
	    \dpp_{{\bf k}_3}\dpp_{{\bf k}_4}\rangle_*^\mathrm{GE} =
	    -\frac{1}{4}(2\pi)^3\delta^3(\sum_a \vect{k}_a)\sum_s
		\ep_{ij}^s(\vect{k}_{12})k_1^ik_2^j
		\ep_{lm}^s(\vect{k}_{34})k_3^l k_4^m\nonumber\\
	 	\times \left[2 \Re \left(
	  	\int_{-\infty}^{\eta_*}\frac{d\eta}{\eta^2}\int_{-\infty}^{\eta}
	    \frac{d\eta'}{\eta'^2}
	    G^{<}_{k_1}(\eta_*,\eta)G^{<}_{k_2}(\eta_*,\eta)
	       G^{<}_{k_3}(\eta_*,\eta')G^{<}_{k_4}(\eta_*,\eta')
	    F^{s<}_{k_{12}}(\eta,\eta') \right) \right. \nonumber\\
	\left. -  \int_{-\infty}^{\eta_*}\frac{d\eta}{\eta^2}\int_{-\infty}^{\eta_*}
	    \frac{d\eta'}{\eta'^2}G^{>}_{k_1}(\eta_*,\eta)G^{>}_{k_2}(\eta_*,\eta)
	       G^{<}_{k_3}(\eta_*,\eta')G^{<}_{k_4}(\eta_*,\eta')
	    F^{s<}_{k_{12}}(\eta,\eta') \right]
	     \nonumber\\
	    \mbox{} + \mbox{23 permutations}\;.
%	    \label{ge_before_simplifications}
	\end{eqnarray}
	Once we explicitly replace the propagators in terms of mode functions,
	this expression 
	yields no factors of $\delta$ in the denominator. In this case, it is
	unnecessary to 
	check that the limit $\delta \rightarrow 0$ is regular and the calculation
	is simplified.
	We would like to thank Peter Adshead, Richard Easther and Eugene Lim for 
	interesting correspondence on this issue. Note, however, that whichever
	method one chooses, the final answer is the same.}

We are therefore left with only the first line of
Eq.~\eref{eq:final_int} which can be computed straightforwardly,
by considering the combined contribution from ${\cal
I}_{1234}$ and ${\cal I}_{3412}$. This yields, using $k_{12}=k_{34}$,
\begin{eqnarray}
 && \fl  {\cal I}_{1234} + {\cal I}_{3412} =
\frac{k_1+k_2}{a_{34}^2}\Bigg[\frac12
(a_{34}+k_{12})(a_{34}^2-2b_{34})
+k_{12}^2 (k_3+k_4) \Bigg] + (1,2 \leftrightarrow 3,4)\nonumber \\
 &&  \!\!\!\!\!\!\!\!\!\!\!\!\!\!\!\!
 +\frac{k_1k_2}{k_t} \Bigg[\frac{b_{34}}{a_{34}} -
k_{12}+ \frac{k_{12}}{a_{12}} \left(k_3k_4 -k_{12}
\frac{b_{34}}{a_{34}}\right)
\left(\frac1{k_t} +\frac1{a_{12}} \right)\Bigg] +(1,2 \leftrightarrow 3,4) \nonumber \\
  &&  \!\!\!\!\!\!\!\!\!\!\!\!\!\!\!\! -
 \frac{ k_{12}}{a_{12} a_{34} k_t} \Bigg[ b_{12} b_{34} + 2
k_{12}^2 \prod_a k_a \left(\frac{1} {k_t^2} +\frac{1}{a_{12} a_{34}}
+\frac{k_{12}}{k_t a_{12} a_{34}} \right) \Bigg]~,
\label{sum_integrals}
\end{eqnarray}
where
$a_{ab}\equiv k_a+k_b+k_{ab}$ and $b_{ab} \equiv (k_a+k_b) k_{ab}
+k_a k_b$.
The first line of Eq.~\eref{sum_integrals} comes from summing all
the contributions which correspond to the first term in the first line
of Eq.~\eref{eq:final_int}, while the last two lines come from
the final three terms.

%%%%%%%%%%%%%%%%%%%%%%%%%%%%%%%%%%%%%%%%%%%%%%%%%%%%%%%%%%%%%%%%%%%%%%%%%%%%
%%%%%%%%%%%%%%%%%%%%%%%%%%%%%%%%%%%%%%%%%%%%%%%%%%%%%%%%%%%%%%%%%%%%%%%%%%%%

\section*{References}

\providecommand{\href}[2]{#2}\begingroup\raggedright\endgroup

\end{fmffile}
\end{document}